# Coupling lattice Boltzmann model for simulation of thermal flows on standard lattices

Q. Li,[1, 2] K. H. Luo,[2] Y. L. He,[1*] Y. J. Gao,[3] and W. Q. Tao[1]

[1]National Key Laboratory of Multiphase Flow in Power Engineering, School of Energy and Power Engineering, Xi'an Jiaotong University, Xi'an, Shaanxi 710049, China

[2]School of Engineering Sciences, University of Southampton, SO17 1BJ Southampton, United Kingdom

[3]Department of Zoology, University of Cambridge, CB2 3EJ Cambridge, United Kingdom

*Corresponding author: yalinghe@mail.xjtu.edu.cn

In this paper, a coupling lattice Boltzmann (LB) model for simulating thermal flows on the standard D2Q9 lattice is developed in the framework of the double-distribution-function (DDF) approach in which the viscous heat dissipation and compression work are considered. In the model, a density distribution function is used to simulate the flow field, while a total energy distribution function is employed to simulate the temperature field. The discrete equilibrium density and total energy distribution functions are obtained from the Hermite expansions of the corresponding continuous equilibrium distribution functions. The pressure given by the equation of state of perfect gases is recovered in the macroscopic momentum and energy equations. The coupling between the momentum and energy transports makes the model applicable for general thermal flows such as non-Boussinesq flows, while the existing DDF LB models on standard lattices are usually limited to Boussinesq flows in which the temperature variation is small. Meanwhile, the simple structure and basic advantages of the DDF LB approach are retained. The model is tested by numerical simulations of thermal Couette flow, attenuation-driven acoustic streaming, and natural convection in a square cavity with small and large temperature differences. The numerical results are found to be in good agreement with the analytical solutions and/or other numerical results reported in the literature.

# Ⅰ. INTRODUCTION

Because of its inherent parallelizability on multiple processors and the avoidance of non-linear convective terms, the lattice Boltzmann (LB) method, which originates from the lattice-gas automata (LGA) method [1], has been developed to be a very attractive numerical method in the past two decades [2-5]. Unlike conventional numerical methods, which are based on the discretization of macroscopic governing equations, the LB method is based on the mesoscopic kinetic equation. The main advantages of the LB method can be summarized as follows [6]: (1) simple form of the governing equation and the easiness of programming: in conventional numerical methods the convection terms of governing equations are non-linear, while in the LB method the convection terms are linear and the viscous effect is modeled through the linearized collision operator, such as the Bhatnagar-Gross-Krook (BGK) collision operator [7-10], Multiple-Relaxation-Time (MRT) collision operator [11-16], and the Two-Relaxation-Time (TRT) collision operator [17, 18]; (2) easy implementation of complex boundary conditions; (3) the LB method is much easier to parallelize and is far less costly in terms of data exchange due to its explicit scheme and the local interaction. It has been demonstrated that the LB method is far better than conventional numerical methods in a parallel implementation [19].

In the LB community, the first LB model for simulating thermal flows was devised by Alexander *et al*. [20], who extended Qian *et al.*'s second-order equilibrium distribution [9] to a third-order model for thermal problems. In the meantime, Qian [21] also proposed a series of higher-order LB models. Since then, several approaches for constructing thermal LB models have been developed, e.g.., the multi-speed approach [20-26], the double-distribution-function (DDF) approach [27-34], and the hybrid approach [35, 36]. The multispeed approach is a straightforward extension of the isothermal LB method. In this approach higher-order velocity moments of the density distribution function are used to describe the energy equation. The DDF approach utilizes two different distribution functions, one for the flow field and the other for the temperature field. In the hybrid approach, the flow simulation is accomplished by the LB method, but the temperature field is solved by conventional numerical methods, such as the finite-difference method.

Despite many efforts have been made from various viewpoints, several limitations still remain in the LB method for thermal flows. The multispeed approach usually suffers from severe numerical instability and the temperature variation simulated is limited to a narrow range [4]. Meanwhile, the boundary treatments become complex when the multispeed lattices are employed: after the streaming process, not only the boundary nodes but also the internal nodes near to the boundaries will possess unknown distribution functions. Lallemand and Luo [35] have found out that the numerical instability of multispeed LB models results from the *spurious* mode coupling in multispeed LB models and cannot be removed by increasing the number of discrete velocities. They suggested that the best approach to remove the *spurious* mode coupling is to treat the energy-conservation equation separately from the mass and momentum conservation equations. Obviously, such a treatment can be realized by the hybrid and the DDF approaches. Nevertheless, as pointed out by Lallemand and Luo [35], the hybrid approach significantly deviates from the orthodox LB method based on the kinetic theory, and it only provides a compromised solution. Meanwhile, the terms of viscous heat dissipation and compressible work are often ignored in the hybrid approach. Fortunately, these terms can be easily considered in the DDF approach. However, most of the existing DDF LB models are "decoupling" models, i.e., an equation of state with a constant temperature ( $p(\rho,T) = \rho c_s^2 = \rho R T_0$, where $c_s = \sqrt{RT_0}$ is the isothermal sound speed, $R$ is the specific gas constant, and $T_0$ is the reference temperature) is recovered. Such a decoupling causes these models to be limited to Boussinesq flows, in which the temperature variation is small. When these models are applied to the thermal problems in which the temperature field has significant influences on the flow field, the decoupling between momentum and energy transports will result in considerable errors.

Recently, Prasianakis *et al.* [37, 38] devised a coupling LB model for simulating thermal flows on standard lattices. Different from previous coupling thermal LB models, Prasianakis *et al.*' model is constructed on the standard D2Q9 lattice. Hence the basic advantages of the standard LB method are retained. Moreover, the Prandtl number is adjustable in the model. However, the model employs a semi-explicit LB scheme. Quasi iterations are therefore introduced in the numerical algorithm of the model. In addition, many gradient terms need to be evaluated in numerical computations. Furthermore, the specific-heat ratio cannot be chosen freely. For the two-dimensional model, the specific-heat ratio is fixed at an unphysical value of 2.

In the present study, we aim to propose a coupling LB model for simulating thermal flows on standard lattices in the framework of the DDF approach on the basis of the following considerations. First, it is well known that in the DDF approach the flow and temperature fields are treated separately, which satisfies the previously mentioned criterion suggested by Lallemand and Luo. Second, the viscous heat dissipation and the compression work can be easily included. Moreover, both the Prandtl number and the specific-heat ratio can be made adjustable in the DDF approach. In the proposed model, the discrete equilibrium distribution functions will be obtained from the Hermite expansions of the continuous equilibrium distribution functions. Particularly, the pressure given by the equation of state of perfect gases will be recovered in the macroscopic momentum and energy equations. The rest of this paper is organized as follows: Section Ⅱ will briefly introduce the DDF LB approach. Section Ⅲ will present the coupling DDF LB model in detail. Both the BGK and MRT versions of the model will be shown. In Sec. IV, numerical simulations will be carried out for several test problems to validate the proposed model. Finally, a brief conclusion is made in Sec. V.

## II. THE DDF LB APPROACH

In this section, we will briefly introduce Guo *et al.*'s DDF LB approach [31]. Historically, the first DDF LB model including the viscous heat dissipation and the compression work was developed by He *et al.* [28]. Two different distribution functions are employed in He *et al.*'s model. The density distribution function is used for the flow field, while an internal energy distribution function is introduced to solve the temperature flied. He *et al.*'s DDF LB model has attracted much attention for its excellent numerical stability over the multispeed LB models and the adjustability of the Prandtl number. Following He *et al.*'s approach, Guo *et al.* [31] developed a total-energy-distribution-function-based DDF LB approach, which enables DDF LB models to be simpler and makes the inclusion of viscous heat dissipation and compression work easier as compared with He *et al.*'s DDF LB approach. The kinetic equations of Guo *et al.*'s DDF LB approach are given as follows [31]:

$$\partial_t f + \boldsymbol{\xi} \cdot \nabla f = -\frac{1}{\tau_f}\left(f - f^{eq}\right), \tag{1a}$$

$$\partial_t h + \boldsymbol{\xi} \cdot \nabla h = -\frac{1}{\tau_h}\left(h - h^{eq}\right) + \frac{Z}{\tau_{hf}}\left(f - f^{eq}\right), \tag{1b}$$

where $f$ and $h$ are the density distribution function and the total energy distribution function, respectively; $\tau_f$ and $\tau_h$ are the momentum and total energy relaxation times, respectively; $\tau_{hf}$ is a relaxation time related to $\tau_f$ and $\tau_h$; and $Z = \boldsymbol{\xi} \cdot \boldsymbol{u} - u^2/2$, where $\boldsymbol{\xi}$ and $\boldsymbol{u}$ are the particle velocity and the macroscopic velocity, respectively. The equilibrium distribution functions $f^{eq}$ and $h^{eq}$ are given by

$$f^{eq} = \frac{\rho}{(2\pi RT)^{D/2}} \exp\left[-\frac{(\boldsymbol{\xi}-\boldsymbol{u})^2}{2RT}\right], \quad h^{eq} = \frac{1}{2}\boldsymbol{\xi}^2 f^{eq} = \frac{\rho \boldsymbol{\xi}^2}{2(2\pi RT)^{D/2}} \exp\left[-\frac{(\boldsymbol{\xi}-\boldsymbol{u})^2}{2RT}\right], \quad (2)$$

where $\boldsymbol{\xi}^2 = \boldsymbol{\xi}\cdot\boldsymbol{\xi}$. By projecting $f^{eq}$ and $h^{eq}$ onto the tensor Hermite polynomial basis in terms of the particle velocity $\boldsymbol{\xi}$, Guo *et al.* obtained the following Hermite expansions of $f^{eq}$ and $h^{eq}$ at the Navier-Stokes level, respectively:

$$f^{eq,3}(T) = \rho\omega(\boldsymbol{\xi},T)\left\{1 + \frac{\boldsymbol{\xi}\cdot\boldsymbol{u}}{RT} + \frac{1}{2}\left(\frac{\boldsymbol{\xi}\cdot\boldsymbol{u}}{RT}\right)^2 - \frac{u^2}{2RT} + \frac{\boldsymbol{\xi}\cdot\boldsymbol{u}}{6RT}\left[\left(\frac{\boldsymbol{\xi}\cdot\boldsymbol{u}}{RT}\right)^2 - \frac{3u^2}{RT}\right]\right\}, \quad (3a)$$

$$h^{eq,2}(T) = \omega(\boldsymbol{\xi},T)\left\{\rho E + (p+\rho E)\frac{\boldsymbol{\xi}\cdot\boldsymbol{u}}{RT} + \frac{p}{2}\left(\frac{\boldsymbol{\xi}^2}{RT} - D\right) + \left(p + \frac{\rho E}{2}\right)\left[\left(\frac{\boldsymbol{\xi}\cdot\boldsymbol{u}}{RT}\right)^2 - \frac{u^2}{RT}\right]\right\}, \quad (3b)$$

where $f^{eq,3}(T)$ is the third-order expansion of $f^{eq}$, $h^{eq,2}(T)$ is the second-order expansion of $h^{eq}$, $E$ is the macroscopic total energy, $p = \rho RT$ is the pressure, and $\omega(\boldsymbol{\xi},T)$ is given by

$$\omega(\boldsymbol{\xi},T) = \frac{1}{(2\pi RT)^{D/2}} \exp\left(-\frac{\boldsymbol{\xi}^2}{2RT}\right). \quad (4)$$

The discrete velocity set $\boldsymbol{e}_\alpha$ can be obtained by choosing the abscissae of a suitable Gauss-Hermite quadrature with the weight function $\omega(\boldsymbol{\xi},T)$. However, as pointed out by Guo *et al.*, the temperature in $\omega(\boldsymbol{\xi},T)$ is a locally changed variable, which means the abscissae of the Gauss-Hermite quadrature are not fixed. Then the discrete velocity $\boldsymbol{e}_\alpha$ will change with the local temperature. To overcome this difficulty, Guo *et al.* replace the local temperature $T$ in $f^{eq,3}(T)$ and $h^{eq,2}(T)$ with a reference temperature $T_0$ [31]:

$$f^{eq,3}(T_0) = \rho\omega(\boldsymbol{\xi},T_0)\left\{1 + \frac{\boldsymbol{\xi}\cdot\boldsymbol{u}}{RT_0} + \frac{1}{2}\left(\frac{\boldsymbol{\xi}\cdot\boldsymbol{u}}{RT_0}\right)^2 - \frac{u^2}{2RT_0} + \frac{\boldsymbol{\xi}\cdot\boldsymbol{u}}{6RT_0}\left[\left(\frac{\boldsymbol{\xi}\cdot\boldsymbol{u}}{RT_0}\right)^2 - \frac{3u^2}{RT_0}\right]\right\}, \quad (5a)$$

$$h^{eq,2}(T_0) = \omega(\boldsymbol{\xi},T_0)\left\{\rho E + (p_0+\rho E)\frac{\boldsymbol{\xi}\cdot\boldsymbol{u}}{RT_0} + \frac{p_0}{2}\left(\frac{\boldsymbol{\xi}^2}{RT_0} - D\right) + \left(p_0 + \frac{\rho E}{2}\right)\left[\left(\frac{\boldsymbol{\xi}\cdot\boldsymbol{u}}{RT_0}\right)^2 - \frac{u^2}{RT_0}\right]\right\}, \quad (5b)$$

where $p_0 = \rho RT_0$. Note that the third-order velocity terms in Eq. (5a) can be neglected in the low Mach number limit.

With regard to the discretization of velocity space, Guo *et al*. adopted a nine-point fifth-degree Gauss-Hermite quadrature, which leads to the following discrete velocity (the D2Q9 lattice):

$$\boldsymbol{e}_\alpha = \begin{cases} (0,0) & \alpha = 0 \\ c\left(\cos[(\alpha-1)\pi/2], \sin[(\alpha-1)\pi/2]\right) & \alpha = 1,2,3,4, \\ \sqrt{2}c\left(\cos[(2\alpha-1)\pi/4], \sin[(2\alpha-1)\pi/4]\right) & \alpha = 5,6,7,8, \end{cases} \quad (6)$$

where $c = \sqrt{3RT_0}$. According to Eq. (1), the kinetic equations for the discrete density distribution function $f_\alpha$ and the discrete total energy distribution function $h_\alpha$ are given by [31]:

$$\partial_t f_\alpha + \boldsymbol{e}_\alpha \cdot \nabla f_\alpha = -\frac{1}{\tau_f}\left(f_\alpha - f_\alpha^{eq}\right), \quad (7a)$$

$$\partial_t h_\alpha + \boldsymbol{e}_\alpha \cdot \nabla h_\alpha = -\frac{1}{\tau_h}\left(h_\alpha - h_\alpha^{eq}\right) + \frac{Z_\alpha}{\tau_{hf}}\left(f_\alpha - f_\alpha^{eq}\right), \quad (7b)$$

where $Z_\alpha = \boldsymbol{e}_\alpha \cdot \boldsymbol{u} - u^2/2$. At the Navier-Stokes level, $Z_\alpha$ can be simplified as $Z_\alpha = \boldsymbol{e}_\alpha \cdot \boldsymbol{u}$. The discrete equilibrium distribution functions $f_\alpha^{eq}$ and $h_\alpha^{eq}$ are given by:

$$f_\alpha^{eq} = \rho w_\alpha \left[1 + \frac{\boldsymbol{e}_\alpha \cdot \boldsymbol{u}}{RT_0} + \frac{1}{2}\left(\frac{\boldsymbol{e}_\alpha \cdot \boldsymbol{u}}{RT_0}\right)^2 - \frac{u^2}{2RT_0}\right], \quad (8a)$$

$$h_\alpha^{eq} = E f_\alpha^{eq} + w_\alpha p_0 \left[\frac{(\boldsymbol{e}_\alpha \cdot \boldsymbol{u})}{RT_0} + \left(\frac{\boldsymbol{e}_\alpha \cdot \boldsymbol{u}}{RT_0}\right)^2 - \frac{u^2}{RT_0} + \frac{1}{2}\left(\frac{e_\alpha^2}{RT_0} - D\right)\right]. \quad (8b)$$

The coefficients are $w_0 = 4/9$, $w_{1,2,3,4} = 1/9$, and $w_{5,6,7,8} = 1/36$. Through the Chapman-Enskog analysis, Guo *et al*. demonstrated that Eq. (7) together with Eq. (8) can recover the following equations in the low Mach number limit:

$$\partial_t \rho + \nabla \cdot (\rho \boldsymbol{u}) = 0, \quad (9a)$$

$$\partial_t (\rho \boldsymbol{u}) + \nabla \cdot (\rho \boldsymbol{u}\boldsymbol{u}) = -\nabla p_0 + \nabla \cdot \boldsymbol{\Pi}, \quad (9b)$$

$$\partial_t (\rho E) + \nabla \cdot \left[(\rho E + p_0)\boldsymbol{u}\right] = \nabla \cdot (\lambda \nabla T) + \nabla \cdot (\boldsymbol{u} \cdot \boldsymbol{\Pi}), \quad (9c)$$

where $\boldsymbol{\Pi} = \mu\left[\nabla \boldsymbol{u} + (\nabla \boldsymbol{u})^T\right]$ is the viscous stress tensor. It can be seen that Eq. (9) is a decoupling Navier-Stokes-Fourier (NSF) equation because the equation of state is given by $p(\rho, T) = p_0 = \rho RT_0$, which makes the momentum transport decoupled from the energy transport. The model is therefore a decoupling model and is limited to Boussinesq flows, in which the temperature variation is small.

## III. THE COUPLING DDF LB MODEL

### A. Hermite expansions of continuous equilibrium distribution functions

From Sec. II it can be seen that the decoupling between the momentum and energy transports in Guo *et al.*'s DDF LB model results from replacing the local temperature $T$ in $f^{eq,3}(T)$ and $h^{eq,2}(T)$ with the reference temperature $T_0$. It is noted that $f^{eq,3}(T)$ given by Eq. (3a) satisfies the zeroth- through third-order velocity moments of the Maxwell distribution function $f^{eq}$:

$$\int f^{eq,3}(T) \mathrm{d}\boldsymbol{\xi} \equiv \int f^{eq} \mathrm{d}\boldsymbol{\xi} = \rho, \tag{10a}$$

$$\int f^{eq,3}(T) \xi_i \mathrm{d}\boldsymbol{\xi} \equiv \int f^{eq} \xi_i \mathrm{d}\boldsymbol{\xi} = \rho u_i, \tag{10b}$$

$$\int f^{eq,3}(T) \xi_i \xi_j \mathrm{d}\boldsymbol{\xi} \equiv \int f^{eq} \xi_i \xi_j \mathrm{d}\boldsymbol{\xi} = \rho u_i u_j + \rho RT \delta_{ij}, \tag{10c}$$

$$\int f^{eq,3}(T) \xi_i \xi_j \xi_k \mathrm{d}\boldsymbol{\xi} \equiv \int f^{eq} \xi_i \xi_j \xi_k \mathrm{d}\boldsymbol{\xi} = \rho u_i u_j u_k + \rho RT \left( u_k \delta_{ij} + u_i \delta_{jk} + u_j \delta_{ik} \right), \tag{10d}$$

where $\delta$ is the Kronecker delta with two indices. However, when the local temperature in $f^{eq,3}(T)$ is replaced with the reference temperature $T_0$, the following relationships will be obtained:

$$\int f^{eq,3}(T_0) \xi_i \xi_j \mathrm{d}\boldsymbol{\xi} = \rho u_i u_j + \rho RT_0 \delta_{ij}, \tag{11a}$$

$$\int f^{eq,3}(T_0) \xi_i \xi_j \xi_k \mathrm{d}\boldsymbol{\xi} = \rho u_i u_j u_k + \rho RT_0 \left( u_k \delta_{ij} + u_i \delta_{jk} + u_j \delta_{ik} \right). \tag{11b}$$

Similar results will be obtained for $h^{eq,2}(T_0)$. As a consequence, the pressure in the macroscopic momentum and energy equations is given by $p(\rho, T) = p_0 = \rho RT_0$ (see Eq. (9)).

In order to overcome the above problem, in the present study we employ another type Hermite expansion of the continuous equilibrium distribution function. As summarized in Refs. [39, 40], in the LB method there exist three different types Hermite expansion of $f^{eq}$. Here we adopt Shan *et al.*'s third-order Hermite expansion of $f^{eq}$, which is given as follows [23]:

$$f^{eq*,3}(T) = \rho\omega(\boldsymbol{\xi},T_0)\left\{1 + \frac{\boldsymbol{\xi}\cdot\boldsymbol{u}}{RT_0} + \frac{1}{2}\left[\left(\frac{\boldsymbol{\xi}\cdot\boldsymbol{u}}{RT_0}\right) - \frac{u^2}{RT_0} + (\theta-1)\left(\frac{\xi^2}{RT_0} - D\right)\right]\right.$$

$$\left. + \frac{\boldsymbol{\xi}\cdot\boldsymbol{u}}{6RT_0}\left[\left(\frac{\boldsymbol{\xi}\cdot\boldsymbol{u}}{RT_0}\right)^2 - \frac{3u^2}{RT_0} + 3(\theta-1)\left(\frac{\xi^2}{RT_0} - D - 2\right)\right]\right\}, \quad (12)$$

where $\theta = T/T_0$. It can be proved that $f^{eq*,3}(T)$ satisfies the zeroth- through third-order velocity moments of $f^{eq}$. In the low Mach number limit, the third-order velocity terms in $f^{eq*,3}(T)$ can be neglected. Then we have

$$f^{eq*,3}(T) = \rho\omega(\boldsymbol{\xi},T_0)\left\{1 + \frac{\boldsymbol{\xi}\cdot\boldsymbol{u}}{RT_0} + \frac{1}{2}\left[\left(\frac{\boldsymbol{\xi}\cdot\boldsymbol{u}}{RT_0}\right) - \frac{u^2}{RT_0} + (\theta-1)\left(\frac{\xi^2}{RT_0} - D\right)\right]\right.$$

$$\left. + \frac{\boldsymbol{\xi}\cdot\boldsymbol{u}}{2RT_0}(\theta-1)\left(\frac{\xi^2}{RT_0} - D - 2\right)\right\}. \quad (13)$$

Inspired by Shan *et al.*' approach for the Hermite expansion of $f^{eq}$ [23], after some algebra we obtain the following second-order Hermite expansion of $h^{eq}$:

$$h^{eq*,2}(T) = \omega(\boldsymbol{\xi},T_0)\left\{\rho E + (p+\rho E)\frac{\boldsymbol{\xi}\cdot\boldsymbol{u}}{RT_0} + \left(p + \frac{\rho E}{2}\right)\left[\left(\frac{\boldsymbol{\xi}\cdot\boldsymbol{u}}{RT_0}\right)^2 - \frac{u^2}{RT_0}\right]\right.$$

$$\left. + \frac{p\theta}{2}\left(\frac{\xi^2}{RT_0} - D\right) + \frac{\rho E}{2}(\theta-1)\left(\frac{\xi^2}{RT_0} - D\right)\right\}. \quad (14)$$

It can be verified that $h^{eq*,2}(T)$ satisfies the zeroth- through second-order velocity moments of $h^{eq}$:

$$\int h^{eq*,2}(T)\mathrm{d}\boldsymbol{\xi} \equiv \int h^{eq}\mathrm{d}\boldsymbol{\xi} = \rho E, \quad (15a)$$

$$\int h^{eq*,2}(T)\xi_i\mathrm{d}\boldsymbol{\xi} \equiv \int h^{eq}\xi_i\mathrm{d}\boldsymbol{\xi} = (\rho E + p)u_i, \quad (15b)$$

$$\int h^{eq*,2}(T)\xi_i\xi_j\mathrm{d}\boldsymbol{\xi} \equiv \int h^{eq}\xi_i\xi_j\mathrm{d}\boldsymbol{\xi} = (\rho E + 2p)u_i u_j + p(E + RT)\delta_{ij}, \quad (15c)$$

where $p = \rho RT$. From Eqs. (5), (12), and (14) we can see that $f^{eq*,3}(T)$ and $h^{eq*,2}(T)$ share a similar feature with $f^{eq,3}(T_0)$ and $h^{eq,2}(T_0)$. Namely, the weight function $\omega(\boldsymbol{\xi},T_0)$ is independent of the local temperature. Hence the discrete velocity $\boldsymbol{e}_\alpha$ will not change with the local temperature. However, as previously mentioned, $f^{eq,3}(T_0)$ and $h^{eq,2}(T_0)$ cannot satisfy the related statistical relationships, which makes the pressure in the macroscopic momentum and energy equations to be given by $p(\rho,T) = \rho RT_0$. Note that $h^{eq*,2}(T)$ can also be written as follows:

$$h^{eq*,2}(T) = \rho E \omega(\boldsymbol{\xi}, T_0)\left\{1 + \frac{\boldsymbol{\xi} \cdot \boldsymbol{u}}{RT_0} + \frac{1}{2}\left[\left(\frac{\boldsymbol{\xi} \cdot \boldsymbol{u}}{RT_0}\right) - \frac{u^2}{RT_0} + (\theta-1)\left(\frac{\boldsymbol{\xi}^2}{RT_0} - D\right)\right]\right\}$$

$$+ \omega(\boldsymbol{\xi}, T_0) p \left[\frac{\boldsymbol{\xi} \cdot \boldsymbol{u}}{RT_0} + \left(\frac{\boldsymbol{\xi} \cdot \boldsymbol{u}}{RT_0}\right)^2 - \frac{u^2}{RT_0} + \frac{\theta}{2}\left(\frac{\boldsymbol{\xi}^2}{RT_0} - D\right)\right]. \quad (16)$$

Therefore, for convenience, we can use the following form of $h^{eq*,2}(T)$:

$$h^{eq*,2}(T) = E f^{eq*,3}(T) + \omega(\boldsymbol{\xi}, T_0) p \left[\frac{\boldsymbol{\xi} \cdot \boldsymbol{u}}{RT_0} + \left(\frac{\boldsymbol{\xi} \cdot \boldsymbol{u}}{RT_0}\right)^2 - \frac{u^2}{RT_0} + \frac{\theta}{2}\left(\frac{\boldsymbol{\xi}^2}{RT_0} - D\right)\right], \quad (17)$$

where $f^{eq*,3}(T)$ is given by Eq. (13).

### B. Discrete equilibrium distribution functions

According to Eqs. (13) and (17), the following discrete equilibrium distribution functions can be obtained when discretizing $\boldsymbol{\xi}$ to $\boldsymbol{e}_\alpha$:

$$f_\alpha^{eq} = \rho w_\alpha \left\{1 + \frac{\boldsymbol{e}_\alpha \cdot \boldsymbol{u}}{RT_0} + \frac{1}{2}\left[\left(\frac{\boldsymbol{e}_\alpha \cdot \boldsymbol{u}}{RT_0}\right)^2 - \frac{u^2}{RT_0} + (\theta-1)\left(\frac{e_\alpha^2}{RT_0} - D\right)\right]\right.$$

$$\left. + \frac{\boldsymbol{e}_\alpha \cdot \boldsymbol{u}}{2RT_0}(\theta-1)\left(\frac{e_\alpha^2}{RT_0} - D - 2\right)\right\}, \quad (18a)$$

$$h_\alpha^{eq} = E f_\alpha^{eq} + w_\alpha p \left[\frac{\boldsymbol{e}_\alpha \cdot \boldsymbol{u}}{RT_0} + \left(\frac{\boldsymbol{e}_\alpha \cdot \boldsymbol{u}}{RT_0}\right)^2 - \frac{u^2}{RT_0} + \frac{\theta}{2}\left(\frac{e_\alpha^2}{RT_0} - D\right)\right]. \quad (18b)$$

Obviously, when $\theta = T/T_0 = 1$ and $p = p_0$, the above discrete equilibrium distribution functions will reduce to those in the decoupling DDF model (see Eq. (8)). When the D2Q9 lattice is adopted, it can be found that $f_\alpha^{eq}$ and $h_\alpha^{eq}$ rigorously satisfy the zeroth- through second-order velocity moments of $f^{eq}$ and $h^{eq}$, respectively, i.e.,

$$\sum_\alpha f_\alpha^{eq} = \rho, \quad \sum_\alpha f_\alpha^{eq} e_{\alpha i} = \rho u_i, \quad \sum_\alpha f_\alpha^{eq} e_{\alpha i} e_{\alpha j} = \rho u_i u_j + \rho RT \delta_{ij}, \quad (19a)$$

$$\sum_\alpha h_\alpha^{eq} = \rho E, \quad \sum_\alpha h_\alpha^{eq} e_{\alpha i} = (\rho E + p) u_i, \quad \sum_\alpha h_\alpha^{eq} e_{\alpha i} e_{\alpha j} = (\rho E + 2p) u_i u_j + p(E + RT) \delta_{ij}. \quad (19b)$$

However, taking the third-order velocity moment of $f_\alpha^{eq}$ with the D2Q9 lattice, we will obtain

$$\sum_\alpha f_\alpha^{eq} e_{\alpha i} e_{\alpha j} e_{\alpha k} = \rho RT \left( u_k \delta_{ij} + u_j \delta_{ik} + u_i \delta_{jk} \right) + 3 \rho RT_0 u_l (1 - \theta) \delta_{ijkl}, \quad (20)$$

where $\delta_{ijkl} = 1$ when $i = j = k = l$, otherwise $\delta_{ijkl} = 0$. Equation (20) can be rewritten as follows:

$$\sum_\alpha f_\alpha^{eq} e_{\alpha i} e_{\alpha j} e_{\alpha k} = \begin{cases} \rho RT_0 \left( u_k \delta_{ij} + u_j \delta_{ik} + u_i \delta_{jk} \right), & \text{if } i = j = k, \\ \rho RT \left( u_k \delta_{ij} + u_j \delta_{ik} + u_i \delta_{jk} \right), & \text{others.} \end{cases} \quad (21)$$

This means that, when the D2Q9 lattice is used, the off-diagonal elements of the third-order velocity moment of $f_\alpha^{eq}$ satisfy the related statistical relationship (see Eq. (10d) and note that the third-order velocity term can be neglected as done in the standard LB method with the low Mach number limit), but the diagonal elements deviate from the relationship. This deviation results from the low symmetry of the D2Q9 lattice and cannot be removed by choosing $f_\alpha^{eq}$ different from Eq. (18a).

For standard lattices, as pointed out by Prasianakis and Karlin [37], the deviation of $f_\alpha^{eq}$ can be removed only by introducing a correction term into the Boltzmann-BGK equation (7a):

$$\partial_t f_\alpha + e_\alpha \cdot \nabla f_\alpha = -\frac{1}{\tau_f} \left( f_\alpha - f_\alpha^{eq} \right) + C_\alpha, \quad (22)$$

where $C_\alpha$ is a correction term. The constraints on $C_\alpha$ can be established with the Chapman-Enskog analysis. By introducing the following multiscale expansions:

$$\partial_t = \varepsilon \partial_{t_1} + \varepsilon^2 \partial_{t_2}, \quad \nabla = \varepsilon \nabla_1, \quad f_\alpha = f_\alpha^{eq} + \varepsilon f_\alpha^{(1)} + \varepsilon f_\alpha^{(2)}, \quad C_\alpha = \varepsilon C_\alpha^{(0)}, \quad (23)$$

where $\varepsilon$ is the expansion parameter, we can rewrite Eq. (22) in the consecutive orders of $\varepsilon$:

$$\varepsilon^1: \left( \partial_{t_1} + e_\alpha \cdot \nabla_1 \right) f_\alpha^{eq} + \frac{1}{\tau_f} f_\alpha^{(1)} = C_\alpha^{(0)}, \quad (24)$$

$$\varepsilon^2: \partial_{t_2} f_\alpha^{(1)} + \left( \partial_{t_1} + e_\alpha \cdot \nabla_1 \right) f_\alpha^{(1)} + \frac{1}{\tau_f} f_\alpha^{(2)} = 0 \quad (25)$$

To recover the correct macroscopic equations, the first two velocity moments of $C_\alpha$ should satisfy

$$\sum_\alpha C_\alpha = 0, \quad \sum_\alpha e_\alpha C_\alpha = 0, \quad (26)$$

Furthermore, taking the second-order velocity moment of Eq. (24) leads to

$$\Pi_{ij} = -\sum_{\alpha} e_{\alpha i} e_{\alpha j} f_{\alpha}^{(1)} = \tau_f \left[ \partial_{t_1} \left( \sum_{\alpha} e_{\alpha i} e_{\alpha j} f_{\alpha}^{eq} \right) + \partial_{1k} \left( \sum_{\alpha} e_{\alpha i} e_{\alpha j} e_{\alpha k} f_{\alpha}^{eq} \right) - \sum_{\alpha} e_{\alpha i} e_{\alpha j} C_{\alpha}^{(0)} \right]. \quad (27)$$

According to Eqs. (20) and (27), we can find that, in order to remove the deviation, the second-order velocity moment of $C_\alpha$ should be defined as

$$\sum_{\alpha} e_{\alpha x} e_{\alpha y} C_{\alpha} = 0, \quad \sum_{\alpha} e_{\alpha x}^2 C_{\alpha} = \partial_x \left[ \rho c^2 (1-\theta) u_x \right], \quad \sum_{\alpha} e_{\alpha y}^2 C_{\alpha} = \partial_y \left[ \rho c^2 (1-\theta) u_y \right]. \quad (28)$$

Here $c = \sqrt{3RT_0} = 1$ is the lattice speed. Equations (26) and (28) are the statistical constraints on the correction term $C_\alpha$. With these constraints, the deviation results from the low symmetry of the D2Q9 lattice can be removed. The detailed form of $C_\alpha$ will be given in the next section.

The full Chapman-Enskog analysis of Eq. (22) and Eq. (7b) with the equilibrium distribution functions given by Eq. (18) is presented in Appendix A. The following macroscopic equations are obtained at the Navier-Stokes level:

$$\partial_t \rho + \nabla \cdot (\rho \boldsymbol{u}) = 0, \quad (29a)$$

$$\partial_t (\rho \boldsymbol{u}) + \nabla \cdot (\rho \boldsymbol{u}\boldsymbol{u}) = -\nabla p + \nabla \cdot \boldsymbol{\Pi}, \quad (29b)$$

$$\partial_t (\rho E) + \nabla \cdot \left[ (\rho E + p) \boldsymbol{u} \right] = \nabla \cdot (\lambda \nabla T) + \nabla \cdot (\boldsymbol{u} \cdot \boldsymbol{\Pi}), \quad (29c)$$

where $\lambda = \tau_h \left[ (b+2)/2 \right] pR$ is the thermal conductivity, in which $b$ is the number of degrees of freedom of the gas. The viscous stress tensor in Eq. (29) is given by

$$\boldsymbol{\Pi} = \mu \left[ \nabla \boldsymbol{u} + (\nabla \boldsymbol{u})^{\mathrm{T}} - \frac{2}{3} (\nabla \cdot \boldsymbol{u}) \mathbf{I} \right] + \mu_{\mathrm{B}} (\nabla \cdot \boldsymbol{u}) \mathbf{I}, \quad (30)$$

where $\mu = \tau_f p$ is the dynamic viscosity, $\mu_{\mathrm{B}} = 2(1/3 - 1/b)\mu$ is the bulk viscosity, and $\mathbf{I}$ is the unit tensor. The Prandtl number of the system is given by $\mathrm{Pr} = \tau_f / \tau_h$. An important difference between Eq. (29) and Eq. (9) can be observed although they take the same form: the pressure in Eqs. (29b) and (29c) is given by $p = \rho RT$, while in Eqs. (9b) and (9c) the pressure is $p_0 = \rho RT_0$.

### C. Time-discrete evolution equations

## 1. BGK scheme

The time discretization of the Boltzmann-BGK equation (22) can be made by integrating the equation along the characteristic line $e_\alpha$ over a time interval of length $\delta_t$, which leads to [28, 31]

$$f_\alpha(x+e_\alpha\delta_t, t+\delta_t) - f_\alpha(x,t) = \int_0^{\delta_t}\left[\Omega_f(x+e_\alpha t', t+t') + C_\alpha(x+e_\alpha t', t+t')\right]dt, \tag{31}$$

where $\Omega_f = (f_\alpha^{eq} - f_\alpha)/\tau_f$. Applying the trapezoidal rule to the integral on the right-hand side yields

$$f_\alpha(x+e_\alpha\delta_t, t+\delta_t) - f_\alpha(x,t) = \frac{\delta_t}{2}\left[\Omega_f(x+e_\alpha\delta_t, t+\delta_t) + \Omega_f(x,t)\right]$$

$$+ \frac{\delta_t}{2}\left[C_\alpha(x+e_\alpha\delta_t, t+\delta_t) + C_\alpha(x,t)\right]. \tag{32}$$

Following He *et al.* [28] and Guo *et al.* [31], the implicitness of Eq. (32) can be eliminated by

$$\bar{f}_\alpha = f_\alpha - \frac{\delta_t}{2}(\Omega_f + C_\alpha). \tag{33}$$

Then Eq. (32) can be rewritten as

$$\bar{f}_\alpha(x+e_\alpha\delta_t, t+\delta_t) - \bar{f}_\alpha(x,t) = -\omega_f\left[\bar{f}_\alpha(x,t) - f_\alpha^{eq}(x,t)\right] - \delta_t(1-0.5\omega_f)C_\alpha(x,t), \tag{34}$$

where $\omega_f = 2\delta_t/(2\tau_f + \delta_t)$. According to Eq. (33), the macroscopic density and velocity can be explicitly computed from $\bar{f}_\alpha$ as follows (note that $\sum_\alpha C_\alpha = \sum_\alpha e_\alpha C_\alpha = 0$):

$$\rho = \sum_\alpha \bar{f}_\alpha, \quad \rho u = \sum_\alpha e_\alpha \bar{f}_\alpha. \tag{35}$$

Similarly, from Eq. (8b), the following time-discrete evolution equation can be obtained:

$$\bar{h}_\alpha(x+e_\alpha\delta_t, t+\delta_t) - \bar{h}_\alpha(x,t) = -\omega_h\left[\bar{h}_\alpha(x,t) - h_\alpha^{eq}(x,t)\right]$$

$$+ (\omega_h - \omega_f)Z_\alpha\left[\bar{f}_\alpha(x,t) - f_\alpha^{eq}(x,t) + \frac{\delta_t}{2}C_\alpha(x,t)\right], \tag{36}$$

where $\omega_h = 2\delta_t/(2\tau_h + \delta_t)$ and $\bar{h}_\alpha = h_\alpha - 0.5\delta_t(\Omega_h - \tau_f Z_\alpha \Omega_f/\tau_{hf})$, in which $\Omega_h = (h_\alpha^{eq} - h_\alpha)/\tau_h$ and $\tau_{hf} = \tau_h\tau_f/(\tau_f - \tau_h)$. The macroscopic total energy can be obtained via $\rho E = \sum_\alpha \bar{h}_\alpha$. According to the definition of the total energy $E = (bRT + u^2)/2$, the macroscopic temperature is given by

$$T = \frac{2}{bR}\left(\sum_\alpha \bar{h}_\alpha/\rho - \frac{1}{2}u^2\right). \tag{37}$$

The specific-heat ratio $\gamma$ is related to $b$ via $\gamma = c_p/c_v = (b+2)/b$, where $c_p = (b+2)R/2$ and $c_v = bR/2$ are the specific heat coefficients at constant pressure and volume, respectively.

For thermal flows with negligible viscous heat dissipation, the second term on the right-hand side of Eq. (36) can be omitted, which leads to

$$\bar{h}_\alpha \left( \bm{x} + \bm{e}_\alpha \delta_t, t + \delta_t \right) - \bar{h}_\alpha \left( \bm{x}, t \right) = -\omega_h \left[ \bar{h}_\alpha \left( \bm{x}, t \right) - h_\alpha^{eq} \left( \bm{x}, t \right) \right]. \tag{38}$$

If the compression work is also negligible, the total energy distribution function can be simplified to an internal energy distribution function with $h_\alpha^{eq} = bRT f_\alpha^{eq}/2$. Then the macroscopic temperature is calculated by $T = 2\sum_\alpha \bar{h}_\alpha / (\rho bR)$.

## 2. MRT scheme

In the LB community, it has been accepted that the MRT collision operator can improve the numerical stability of the LB equation as compared with the BGK collision operator. Because the relaxation times for hydrodynamic and non-hydrodynamic moments can be separated when employing the MRT collision operator [11-16]. By replacing the BGK collision operator with the MRT collision operator, Eq. (22) can be rewritten as

$$\partial_t f_\alpha + \bm{e}_\alpha \cdot \nabla f_\alpha = -\Lambda_{\alpha\beta} \left( f_\beta - f_\beta^{eq} \right) + C_\alpha, \tag{39}$$

where $\Lambda = \bm{M}^{-1}\bm{S}\bm{M}$ is the collision matrix, $\bm{M}$ is an orthogonal transformation matrix (see Appendix B), and $\bm{S}$ is a diagonal Matrix given by

$$\bm{S} = \mathrm{diag}\left( \tau_\rho^{-1}, \tau_e^{-1}, \tau_\varsigma^{-1}, \tau_j^{-1}, \tau_q^{-1}, \tau_j^{-1}, \tau_q^{-1}, \tau_f^{-1}, \tau_f^{-1} \right). \tag{40}$$

Through the transformation matrix, the density distribution function $f_\alpha$ can be projected onto the moment space with $\bm{m} = \bm{M}\bm{f}$, where $\bm{f} = \left( f_0, f_1, \cdots, f_8 \right)^{\mathrm{T}}$. For the D2Q9 lattice, $\bm{m}$ is defined as $\bm{m} = \left( \rho, e, \varsigma, j_x, q_x, j_y, q_y, p_{xx}, p_{xy} \right)^{\mathrm{T}}$, where $e$ is the energy mode, $\varsigma$ is related to energy square, $(j_x, j_y)$ are the momentum components, $(q_x, q_y)$ correspond to energy flux; and $(p_{xx}, p_{xy})$ are related to the diagonal and off-diagonal components of the stress tensors [12]. Similarly, $f_\alpha^{eq}$ can be projected onto the moment space with $\mathbf{m}^{eq} = \mathbf{M}\mathbf{f}^{eq}$, where $\mathbf{f}^{eq} = \left(f_0^{eq}, \cdots, f_8^{eq}\right)^{\mathrm{T}}$. According to Eq. (18a), after some algebra we can obtain the equilibria $\mathbf{m}^{eq}$ of the present model as follows:

$$\begin{aligned}\mathbf{m}^{eq} &= \left(\rho, e^{eq}, \varsigma^{eq}, j_x, q_x^{eq}, j_y, q_y^{eq}, p_{xx}^{eq}, p_{xy}^{eq}\right)^{\mathrm{T}} \\ &= \rho\Big[1, \left(-4+3\boldsymbol{u}^2+2\theta\right), \left(3-3\boldsymbol{u}^2-2\theta\right), u_x, \\ &\quad \left(-2u_x+u_x\theta\right), u_y, \left(-2u_y+u_y\theta\right), u_x^2-u_y^2, u_x u_y\Big]^{\mathrm{T}},\end{aligned} \quad (41)$$

When $\theta = T/T_0 = 1$, the equilibria will reduce to those in the isothermal D2Q9 MRT model [42, 43]. It should be noted that the moment corresponding to the energy, $m_2 = e$, is not a conserved moment. The energy conservation is separately descried by the total energy distribution function.

The time discretization of Eq. (39) can also be made by integrating the equation along the characteristic line. Nevertheless, the MRT collision term needs to be integrated explicitly. Then the following equation will be obtained:

$$f_\alpha\left(\boldsymbol{x}+\boldsymbol{e}_\alpha\delta_t, t+\delta_t\right) - f_\alpha(\boldsymbol{x},t) = -\delta_t\Lambda_{\alpha\beta}\left(f_\beta - f_\beta^{eq}\right)\Big|_{(\boldsymbol{x},t)} + \frac{\delta_t}{2}\left(C_\alpha\big|_{(\boldsymbol{x}+\boldsymbol{e}_\alpha\delta_t,t+\delta_t)} + C_\alpha\big|_{(\boldsymbol{x},t)}\right). \quad (42)$$

By using $\hat{f}_\alpha = f_\alpha - 0.5\delta_t C_\alpha$, we can rewrite Eq. (42) as follows:

$$\hat{f}_\alpha\left(\boldsymbol{x}+\boldsymbol{e}_\alpha\delta_t, t+\delta_t\right) = \hat{f}_\alpha(\boldsymbol{x},t) - \delta_t\Lambda_{\alpha\beta}\left(\hat{f}_\beta - f_\beta^{eq}\right)\Big|_{(\boldsymbol{x},t)} + \delta_t\left(C_\alpha - 0.5\Lambda_{\alpha\beta}C_\beta\right)\Big|_{(\boldsymbol{x},t)}. \quad (43)$$

The right-hand side of Eq. (43) can be carried out in the moment space

$$\hat{\mathbf{m}}^* = \hat{\mathbf{m}} - \delta_t \mathbf{S}\left(\hat{\mathbf{m}}-\mathbf{m}^{eq}\right) + \delta_t\left(\mathbf{I}-\frac{\mathbf{S}}{2}\right)\widehat{\mathbf{C}}, \quad (44)$$

where $\hat{\mathbf{m}} = \mathbf{M}\hat{\mathbf{f}}$ and $\widehat{\mathbf{C}} = \mathbf{M}\mathbf{C}$, in which $\mathbf{C} = \left(C_0, C_1, \cdots, C_8\right)^{\mathrm{T}}$. The streaming process is given by

$$\hat{f}_\alpha\left(\boldsymbol{x}+\boldsymbol{e}_\alpha\delta_t, t+\delta_t\right) = \hat{f}_\alpha^*(\boldsymbol{x},t), \quad (45)$$

where $\hat{\mathbf{f}}^* = \mathbf{M}^{-1}\hat{\mathbf{m}}^*$. The macroscopic density and velocity are calculated by

$$\rho = \sum_\alpha \hat{f}_\alpha, \quad \rho\boldsymbol{u} = \sum_\alpha \boldsymbol{e}_\alpha \hat{f}_\alpha. \quad (46)$$

With $\bar{f}_\alpha = f_\alpha - 0.5\delta_t C_\alpha$, the evolution equation for the total energy distribution function is given by

$$\bar{h}_\alpha\left(\boldsymbol{x}+\boldsymbol{e}_\alpha\delta_t, t+\delta_t\right) - \bar{h}_\alpha(\boldsymbol{x},t) = -\omega_h\left[\bar{h}_\alpha(\boldsymbol{x},t) - h_\alpha^{eq}(\boldsymbol{x},t)\right]$$

$$+\left(1-0.5\omega_h\right)\frac{\delta_t Z_\alpha}{\tau_{hf}}\left[\hat{f}_\alpha\left(\boldsymbol{x},t\right)-f_\alpha^{eq}\left(\boldsymbol{x},t\right)+\frac{\delta_t}{2}C_\alpha\left(\boldsymbol{x},t\right)\right]. \qquad (47)$$

According to the Chapman-Enskog analysis, the shear viscosity in the energy equation (contained in $\Pi$) will be given by $\mu = \left(\tau_f \tau_h / \tau_{hf} + \tau_h\right)p$, while the shear viscosity in the macroscopic momentum equation from Eq. (43) is $\mu = \left(\tau_f - 0.5\delta_t\right)p$. To ensure these two viscosities are the same, the relaxation time $\tau_{hf}$ should be defined as $\tau_{hf} = \tau_f/(\text{Pr}-1)$. The Prandtl number is now given by $\text{Pr} = \mu c_p/\lambda = \left(\tau_f - 0.5\delta_t\right)/\tau_h$.

Furthermore, theoretically speaking, the relaxation time $\tau_e$ in Eq. (40) should be equal to $\tau_f$ to ensure that the bulk viscosity $\mu_B$ in the energy equation is consistent with that in the momentum equation. Actually, if $\tau_e = \tau_\varsigma = \tau_f$, the MRT collision operator will become a TRT collision operator, which is also better than the BGK collision operator in terms of numerical stability [14]. In practical applications, different values of $\tau_e$ and $\tau_f$ can be adopted since the influence of the bulk viscosity on the energy equation can be neglected in most of low Mach number flows.

### 3. Correction term

Now we give the detailed form of the correction term. Actually, in the moment space the correction term $\hat{\mathbf{C}}$ can be directly determined via the Chapman-Enskog analysis. Details about the Chapman-Enskog analysis of isothermal MRT models can be found in Refs. [40, 41]. The Chapman-Enskog analysis of the present MRT model can be conducted in a similar way, and we can obtain:

$$\hat{\mathbf{C}} = \left(0, \varphi_x, 0, 0, 0, 0, 0, \varphi_y, 0\right)^{\text{T}}, \quad \varphi_x = 3\left(\partial_x Q_x + \partial_y Q_y\right), \quad \varphi_y = \left(\partial_x Q_x - \partial_y Q_y\right), \qquad (48)$$

where $Q_x = \rho u_x(1-\theta)$ and $Q_y = \rho u_y(1-\theta)$. $Q_x$ and $Q_y$ are the order of $\text{Ma}\Delta T$, where Ma is the local Mach number and $\Delta T = (T-T_0)/T_0$. The correction term in the velocity space, $C_\alpha$, can be obtained via the relation $\mathbf{C} = \mathbf{M}^{-1}\hat{\mathbf{C}}$. The transformation matrix $\mathbf{M}$ of the D2Q9 lattice and its inverse matrix $\mathbf{M}^{-1}$ are given in Appendix B. According to Eq. (48) and Eq. (B2), $C_\alpha$ is given by

$$C_0 = -\frac{1}{9}\varphi_x, \quad C_1 = -\frac{1}{36}\varphi_x + \frac{1}{4}\varphi_y, \quad C_2 = -\frac{1}{36}\varphi_x - \frac{1}{4}\varphi_y,$$

$$C_3 = C_1, \quad C_4 = C_2, \quad C_5 = C_6 = C_7 = C_8 = \frac{1}{18}\varphi_x. \tag{49}$$

It can be easily verified that $C_\alpha$ given by Eq. (49) satisfies the constraints in Eqs. (26) and (28). In numerical simulations, the gradient terms $\partial_x Q_x$ and $\partial_y Q_y$ in Eq. (48) can be evaluated as follows:

$$\partial_x Q_x\big|_{(I,J)} = \frac{1}{2\delta_x}\Big[Q_x\big|_{(I+1,J)} - Q_x\big|_{(I-1,J)}\Big], \quad \partial_y Q_y\big|_{(I,J)} = \frac{1}{2\delta_y}\Big[Q_y\big|_{(I,J+1)} - Q_y\big|_{(I,J-1)}\Big], \tag{50}$$

where $\delta_x$ and $\delta_y$ are the spatial steps in the $x$- and $y$-directions, respectively. If needed, one can also adopt some other schemes [42, 43]. In fact, for low Mach number flows with small/moderate temperature variations, the correction term is usually negligible.

When a force $\mathbf{a}$ is exerted on the flow, a forcing term $F_\alpha$ should be included in Eqs. (22) and (39). $F_\alpha$ can be defined as $F_\alpha = w_\alpha\Big[(\mathbf{e}_\alpha \cdot \mathbf{a})/(RT_0) + \mathbf{ua}:(\mathbf{e}_\alpha\mathbf{e}_\alpha - RT_0\mathbf{I})/(RT_0)^2\Big]$ [31]. The forcing term in the moment space can be obtained via $\hat{\mathbf{F}} = \mathbf{MF}$, i.e.,

$$\hat{\mathbf{F}} = \Big[0,\ 6\mathbf{u}\cdot\mathbf{a},\ -6\mathbf{u}\cdot\mathbf{a},\ a_x,\ -a_x,\ a_y,\ -a_y,\ 2(u_x a_x - u_x a_y),\ (u_x a_y + u_y a_x)\Big]^T. \tag{51}$$

In the evolution equations, $F_\alpha$ and $\hat{\mathbf{F}}$ can be absorbed into $C_\alpha$ and $\hat{\mathbf{C}}$, respectively. When the forcing term is included, the macroscopic velocity will be given by $\rho\mathbf{u} = \sum_\alpha \mathbf{e}_\alpha \hat{f}_\alpha + 0.5\delta_t \mathbf{a}$.

Finally, a brief comparison between the present model and Prasianakis *et al.*'s coupling BGK-LB model [37, 38] is made. First, it is noted that these two models share some similar features: both models are based on the standard D2Q9 lattice; the Prandtl number is adjustable; and a correction term is introduced into the kinetic equation due to the low symmetry of the D2Q9 lattice. On the other hand, the differences between the two models are also apparent. The present model is constructed in the framework of the DDF approach. Hence only the deviation of the third-order velocity moment of $f_\alpha^{eq}$ needs to be considered, while Prasianakis *et al.*' model is based on the single density distribution function and should consider the deviations of the third-order and fourth-order velocity moments of $f_\alpha^{eq}$. As a result, more than 10 gradient terms need to be computed in Prasianakis *et al.*'s model.

Moreover, from Eq. (27) it can be seen that, in the present model, the deviation of the third-order velocity moment of $f_\alpha^{eq}$ is removed via the second-order velocity moment of the correction term. Such a treatment is consistent with the philosophy of the Chapman-Enskog analysis. However, in Prasianakis *et al.*'s model, the related deviation is removed by the first-order velocity moment of the correction term, which makes the computational scheme of the macroscopic velocity $\boldsymbol{u}$ implicit. Quasi iterations are therefore introduced in the numerical algorithm of Prasianakis *et al.*'s model [37]. In addition, the specific-heat ratio $\gamma$, which is fixed at an unphysical value of $(D+2)/D$ in Prasianakis *et al.*'s model, can be freely chosen in the present model.

## IV. NUMERICAL TESTS

In this section, we conduct a series of numerical simulations to validate the proposed coupling DDF LB model. Without loss of generality, the MRT version of the model is used in the simulations. The testing problems include thermal Couette flow, attenuation-driven acoustic streaming, and natural convection in a square cavity.

### A. Thermal Couette flow

Thermal Couette flow is a classical heat transfer problem which can provide a good test of new LB models to describe the viscous heat dissipation. The existing LB simulations of Couette flows are mostly conducted under the assumption of constant viscosity and thermal conductivity coefficients. In the present work, we pay attention to thermal Couette flows with temperature-dependent transport coefficients, in which the shear viscosity and the thermal conductivity vary with the temperature through a power law or the Sutherland law or some other empirical functions. Since the viscosity depends on the temperature, the velocity and temperature fields are coupled together.

Consider the viscous fluid flow between two infinite parallel plates, the lower plate is fixed and the upper one is moving at speed $U$. The ratio $\lambda/\mu = c_p/\text{Pr}$ is assumed to be constant, and this is reasonable for gases [44]. In fact, there are no closed-form solutions for a general viscosity model. However, under the conditions of $\mu/\mu_0 = (T/T_0)^n$ with $n=1$ and of adiabatic lower wall, there will be an analytical solution as follows [44]:

$$\frac{T}{T_0} = 1 + \Pr\frac{\gamma-1}{2}\text{Ma}^2\left[1-\left(\frac{u_x}{U}\right)^2\right], \tag{52a}$$

$$\left(1+\Pr\frac{\gamma-1}{3}\text{Ma}^2\right)\frac{y}{H} = \frac{u_x}{U} + \Pr\frac{\gamma-1}{2}\text{Ma}^2\left[\frac{u_x}{U}-\frac{1}{3}\left(\frac{u_x}{U}\right)^3\right], \tag{52b}$$

where $T_0$ is the temperature of the upper wall and $H$ is the distance between the two plates. The Mach number is defined as $\text{Ma} = U/\sqrt{\gamma R T_0}$.

Numerical simulations are carried out with different values of Pr and Ma: (1) $\Pr = 1$, 3, and 5 with $\gamma = 5/3$ ($b=3$) and $\text{Ma} = 0.35$; (2) $\text{Ma} = 0.15$, 0.25, and 0.35 (in the low Mach number regime) with $\gamma = 5/3$ and $\Pr = 5$. A grid size of $N_x \times N_y = 10 \times 60$ ($N_x$ and $N_y$ exclude the extra layers outside the boundaries) is employed, and the non-equilibrium extrapolation method [31, 45] is used to treat the velocity and temperature boundary conditions of the upper and lower walls, while the periodic boundary condition is applied in the *x*-direction.

In simulations, we set $\rho_0 = 1$, $T_0 = 1$, $\mu_0 = 0.25$, $c = \sqrt{3RT_0} = 1$, and $\delta_x = \delta_y = \delta_t = 1$. The relaxation times are chosen as follows: $\tau_e^{-1} = \tau_\varsigma^{-1} = \tau_q^{-1} = 1.5$ and $\tau_\rho = \tau_j = 1$. From Eq. (52a) it can be seen that $(\theta - 1) = (T/T_0 - 1)$ is the order of $\text{Ma}^2$. Therefore the correction term is omitted since it will be the order of $Ma^3$. The temperature profiles are shown in Figs. 1 and 2, in which the analytical results are also presented for comparison. It is clearly seen that the numerical results are in good agreement with the analytical ones. To quantify the results, the numerical ($T$) and analytical ($T_a$) temperatures at the lower wall are listed in Table I. The relative error defined by $E_r = |(\theta-1)-(\theta_a-1)|/(\theta_a-1)$ is computed, where $\theta_a = T_a/T_0$. As shown in Table I, the difference between the present results and the analytical results are within 1.0% for the cases considered.

### B. Attenuation-driven acoustic streaming

Now we consider an acoustic problem, the acoustic streaming, which describes a flow field superimposed upon the oscillatory motion of a sound wave propagating in a fluid. It is a non-linear effect which occurs due to the presence of boundaries or because of damping of waves [46]. Acoustic streaming has important and possibly undesirable effects in the study of acoustic levitation and the operation of thermo-acoustic systems. There are two basic types of acoustic streaming. One is the Rayleigh streaming, and the other is the attenuation-driven acoustic streaming, which results from the attenuation of the wave in a fluid.

In this test, the attenuation-driven acoustic streaming is simulated. A plane acoustic wave of wave-vector $k = 2\pi/\lambda'$ and angular frequency $\omega = kc_s$ ($c_s = \sqrt{\gamma RT}$ is the speed of sound) propagates in the positive *x*-direction is considered. The wave propagates from a source at $x = 0$. The problem is initialized with a zero velocity and undisturbed density and temperature everywhere. The sound wave is generated by adding sinusoidal density and temperature variations

$$\rho = \rho_0 + \delta\rho \sin(\omega t), \quad T = T_0 + (\gamma-1)\frac{\delta\rho}{\rho_0}T_0 \sin(\omega t), \quad (53)$$

at $x = 0$. Then the analytical velocity is given by [47]

$$u_x(x,t) = U e^{-\beta x} \sin(\omega t - kx), \tag{54}$$

where $U = c_s \delta \rho / \rho_0$ and $\beta$ is the absorption coefficient. The velocity amplitude is given by

$$U(x) = U e^{-\beta x}. \tag{55}$$

The absorption coefficient $\beta$ can be obtained via the perturbation solution of the NSF equations [47]:

$$\beta = \frac{2\pi^2 f'^2}{\rho c_s^3} \left[ \frac{4}{3}\mu + \mu_B + \left(\frac{1}{c_v} - \frac{1}{c_p}\right)\lambda \right], \tag{56}$$

where $f'$ is the frequency. Since Eq. (56) is based on a perturbation solution, the density perturbation $\delta\rho$ should be small enough to ensure Eq. (56) is valid. In the present study, we set $\delta\rho = 0.01\rho_0$.

In simulations, we choose the region $x \in [0, 500]$ as the measurement region. The length along $x$-direction, $L_x = N_x \times \delta_t$, where $N_x$ is the lattice number in the $x$-direction, should be large enough so that the density disturbance will not reach $x = L_x$ during the measurement procedure, namely $L_x > c_s t_m > 500$, where $t_m$ is the measurement time. In the computation, we set $\rho_0 = 1$, $T_0 = 1$, $\mu = 0.1$, $\lambda' = 50$, $c = 1$, $\delta_x = \delta_y = \delta_t = 1$, $\tau_e = \tau_\varsigma = \tau_f$, $\tau_q^{-1} = 1.5$, $\tau_\rho = \tau_j = 1$, $t_m = 800$, and $L_x = 600$. A grid size of $N_x \times N_y = 600 \times 4$ is employed. The inlet boundary condition is implemented by constantly resetting the equilibrium distributions with the desired values of density, velocity, and temperature. At the outlet, the zero-gradient condition is imposed by simply extrapolating the information ($\rho, \mathbf{u}, T$) from the fluid field near the outlet boundary, and then the outlet boundary condition can be implemented by using the non-equilibrium extrapolation method. The periodic boundary condition is applied in the $y$-directions. The velocity profiles at $\gamma = 5/3$ with $\text{Pr} = 0.2$, $0.5$, and $5$ are presented in Fig. 3, and the velocity profiles at $\text{Pr} = 2$ with $\gamma = 5/3$, $7/5$, and $9/7$ are shown in Fig. 4. For comparison, the analytical velocity amplitudes are also shown in Figs. 3 and 4. It can be seen that the numerical results agree well with the analytical solutions.

A comparison of the numerical results predicted by the present model and the decoupling model

($\theta=1$) is presented in Fig. 5, from which we can observe that the numerical results predicted by the decoupling model significantly deviate from the analytical ones. Actually, in this test the correction term is found to be negligible. Hence we have omitted it. Then the only difference between the present model and the decoupling model lies in that the terms related to $\theta$ are included in the equilibrium distribution functions of the present model. From Eq. (53) it can be derived that the maximum value of $\theta$ is $\theta_{max} = 1+(\gamma-1)\delta\rho/\rho_0$, which is approximately equal to 1 since $\delta\rho/\rho_0 = 0.01$. However, Fig. 5 clearly illustrates that directly setting $\theta=1$ will result in considerable errors.

In fact, when $\theta$ is set to be unity, the momentum and energy transports will be decoupled. As a result, the term related to the thermal conductivity $\lambda$ will vanish in Eq. (56). In addition, the sound speed $c_s$ will be given by $c_s = \sqrt{RT_0}$ rather than $c_s = \sqrt{\gamma RT_0}$. Then the efficient wavelength should be defined as $\lambda'_{eff} = \lambda'/\sqrt{\gamma}$. In other words, the wavelength predicted by the decoupling model will be smaller than the one predicted by the coupling model, which can be clearly seen in Fig. 5. This test demonstrates that in acoustic problems the temperature field will have significant influences on the flow field even though the temperature variation is very small.

### C. Natural convection in a square cavity

Natural convection in a square cavity with adiabatic top and bottom walls and with side walls maintained at constant but different temperatures is often taken as one of the standard cases to test new computational schemes. This flow is characterized by two non-dimensional parameters: the Prandtl number Pr and the Rayleigh number Ra, which is defined as

$$\text{Ra} = \frac{g\beta(T_h - T_l)H^3 \text{Pr}}{\mu_r^2}, \tag{57}$$

where $g$ is the gravity acceleration, $\beta$ is the thermal expansion coefficient, $T_h$ and $T_l$ are the temperatures of left and right walls ($T_h > T_l$), $H$ is the distance between the walls, and $\mu_r$ is the shear viscosity evaluated at the reference temperature $T_r = (T_h + T_l)/2$. For perfect gases, the thermal expansion coefficient $\beta$ is defined as

$$\beta = -\frac{1}{\rho}\left(\frac{\partial \rho}{\partial T}\right)_P = \frac{1}{T}. \tag{58}$$

The thermal expansion coefficient in Eq. (57) is also evaluated at $T_r$.

Firstly, we simulate the natural convection with a small temperature difference: $T_h$ and $T_l$ are set to be $315\,\mathrm{K}$ and $285\,\mathrm{K}$, respectively. With such a temperature difference, the Boussinesq assumption is approximately satisfied. However, since the present model is a coupling model, we do not need to use the Boussinesq assumption. We can directly set the force $\boldsymbol{a}$ in Eq. (51) as follows: $a_x = 0$ and $a_y = -\rho g$. The viscous heat dissipation and the compression work can be neglected. Hence Eq. (38) with the internal energy distribution function is adopted. The gradient terms in the correction term are evaluated by Eq. (50) in this test (including the case with a large temperature difference given below). The relaxation times are chosen as follows: $\tau_e = \tau_\varsigma = \tau_f$, $\tau_q^{-1} = 1.5$, and $\tau_\rho = \tau_j = 1$. The Prandtl number is set to be $\mathrm{Pr} = 0.71$. The temperature $T_0$ in the lattice speed can be chosen as $T_0 = T_r$, and then $R$ is determined by $R = c^2/(3T_0)$. The non-equilibrium extrapolation scheme [31, 45] is employed to treat different boundary conditions of $\widehat{f}_\alpha$ and $\overline{h}_\alpha$.

Through the grid-dependence examination of the numerical results, the grid sizes of $100 \times 100$ for $\mathrm{Ra} = 10^3$, $150 \times 150$ for $\mathrm{Ra} = 10^4$, and $200 \times 200$ for $10^5$ are adopted. The streamlines and isotherms are shown in Fig. 6. From the figure it can be observed that for low values of Rayleigh number, a vortex appears in the center of the cavity. The vortex gradually becomes elliptic as the Rayleigh number increases, and breaks up into two vortices at $\mathrm{Ra} = 10^5$. Meanwhile, the isotherms clearly show that, for the case of $\mathrm{Ra} = 10^3$, heat is transferred mainly by conduction between the hot and cold walls. Hence the isotherms are almost vertical. When the Rayleigh number increases, the dominant heat transfer mechanism changes from conduction to convection, and then the isotherms become horizontal in the center of the cavity. These observations are in good agreement with the results reported in the literature [31, 48].

To quantify the results, the maximum horizontal velocity component $u_{max}$ in the vertical midplane $x = H/2$ and its location $y_{max}$, the maximum vertical component $v_{max}$ in the horizontal midplane $y = H/2$ and its location $x_{max}$; and the average Nusselt number $Nu$ along the cold wall are computed. The $x$- and $y$-coordinates are normalized with the cavity width $H$. The velocities are normalized with the diffusion velocity $V_{diff} = \chi/H$, where $\chi = \tau_h p$ is the thermal diffusivity. The results are listed in Table II together with the data from previous studies. As shown, the present results agree well with the available data.

Furthermore, the case of natural convection with a large temperature difference is also considered: $T_h = 960\,\text{K}$ and $T_l = 240\,\text{K}$. The shear viscosity is defined by the Sutherland's law:

$$\frac{\mu(T)}{\mu^*} = \left(\frac{T}{T^*}\right)^{\frac{3}{2}} \frac{(T^* + S)}{(T + S)}, \qquad (59)$$

where $T^* = 273\,\text{K}$, $S = 110.5\,\text{K}$, and $\mu^*$ is the shear viscosity evaluated at $T^*$. The Prandtl number is assumed to remain constant, i.e., $Pr = 0.71$, and this is reasonable for gases. For example, for air from 0 to 1000° F, the ratio $\lambda/\mu$ increases only 20 percent [44].

In simulations, the relaxation time $\tau_f$ is determined by $\tau_f = \mu/p + 0.5\delta_t$, where $\mu$ is calculated by Eq. (59). The viscosity $\mu_r$ in Eq. (57) is set to be $\mu_r = 0.1$, 0.1, and 0.06 for $Ra = 10^3$, $10^4$, and $10^5$, respectively. The viscosity $\mu^*$ in Eq. (59) is determined from $\mu(T_r)$. The grid sizes of $100 \times 100$, $150 \times 150$, and $200 \times 200$ are adopted for $Ra = 10^3$, $10^4$, and $10^5$, respectively. The streamlines and isotherms are presented in Figs. 7 and 8. For comparison, the numerical results at $Ra = 10^5$ with the same temperatures reported in Ref. [49] are also shown in Fig. 8, from which good agreement can be observed. The asymmetry feature of natural convection with a large temperature difference can be clearly seen in Figs. 7 and 8: there is a very apparent shift of the centre of the primary vortex both towards the right wall and downwards towards the bottom of the cavity.

By comparing the present case with the case of natural convection with a small temperature difference, we can find that the temperature gradients become higher near the right wall but lower near the left wall. In fact, this is because that the local Rayleigh number near the right wall is larger than the defined Rayleigh number, while the local Rayleigh number near the left wall is smaller than the defined Rayleigh number. For $Ra = 10^5$, the maximum Rayleigh number near the right wall is around $9.5 \times 10^5 \approx 10^6$, whereas the minimum Rayleigh number near the left wall is only about $0.35 \times 10^5$. Figure 9 illustrates the temperature profiles along the horizontal line crossing the center of the cavity. From the figure the symmetry and asymmetry features of natural convection in different cases can be clearly observed.

## Ⅴ. CONCLUSION

In this paper, a coupling LB model has been developed for simulating thermal flows on the standard D2Q9 lattice in the framework of the DDF LB approach. In the model, the discrete equilibrium distribution functions are derived from the Hermite expansions of the continuous equilibrium distribution functions. When the D2Q9 lattice is used, these discrete equilibrium distribution functions can rigorously satisfy the zeroth- through second-order velocity moments of the corresponding continuous equilibrium distribution functions. However, the third-order velocity moment of the discrete equilibrium density distribution function deviates from the related statistical relationship. To remove this deviation, a correction term has been introduced into the kinetic equation. The statistical constraints on the correction term and the detailed form of the correction term have been determined with the Chapman-Enskog analysis. The BGK and MRT versions of the present model have been both proposed.

Numerical simulations have been performed for thermal Couette flow with temperature-dependent transport coefficients, attenuation-driven acoustic streaming with different Prandtl numbers and specific-heat ratios, and natural convection in a square cavity with small and large temperature differences. The numerical experiments show that the results predicted by the present model are in good agreement with the analytical solutions and/or the numerical results reported in the literature.

In summary, the present model exhibits some features that distinguish it from the existing LB models. First, unlike previous DDF LB models based on standard lattices, the present model can ensure the momentum and energy transports are physically coupled, which makes it applicable for general thermal flows such as non-Boussinesq flows. Second, different from most of the existing coupling LB models which are usually constructed on multispeed lattices, the present model is based on the DDF approach with the standard D2Q9 lattice. The simple structure and the basic advantages of the DDF approach have been retained in the model. Furthermore, compared with Prasianakis *et al.*'s coupling BGK-LB model which is also based on the standard D2Q9 lattice, the present model has the following advantages: the correction term is much simpler, the computational scheme of the macroscopic velocity is explicit, and the specific-heat ratio is adjustable. With the formulations in the present paper, the model can be readily extended to three-dimensional (3D) space and other systems such as flows of non-Newtonian fluids [50] and axisymmetric flows in cylindrical coordinates [15].

**ACKNOWLEDGMENTS**

This work was supported by the National Natural Science Foundation of China (No. 50736005 and No. 51176155) and the Engineering and Physical Sciences Research Council of United Kingdom under Grant Nos. EP/I000801/1 and EP/I012605/1.

**APPENDIX A: CHAPMAN-ENSKOG ANALYSIS**

In this Appendix, the Chapman-Enskog analysis of Eq. (22) and Eq. (7b) with the equilibrium distribution functions in Eq. (18) is presented. Substituting the expansions given by Eq. (23) and the expansion $h_\alpha = h_\alpha^{eq} + \varepsilon h_\alpha^{(1)} + \varepsilon h_\alpha^{(2)}$ into Eq. (22) and Eq. (7b), we can obtain a series of equations in terms of the order of $\varepsilon$:

$$\varepsilon^1 : \left(\partial_{t_1} + \boldsymbol{e}_\alpha \cdot \boldsymbol{\nabla}_1\right) f_\alpha^{eq} + \frac{1}{\tau_f} f_\alpha^{(1)} = C_\alpha^{(0)}, \tag{A1}$$

$$\varepsilon^2 : \partial_{t_2} f_\alpha^{(1)} + \left(\partial_{t_1} + \boldsymbol{e}_\alpha \cdot \boldsymbol{\nabla}_1\right) f_\alpha^{(1)} + \frac{1}{\tau_f} f_\alpha^{(2)} = 0, \tag{A2}$$

and

$$\varepsilon^1 : \left(\partial_{t_1} + \boldsymbol{e}_\alpha \cdot \boldsymbol{\nabla}_1\right) h_\alpha^{eq} + \frac{h_\alpha^{(1)}}{\tau_h} = Z_\alpha \frac{f_\alpha^{(1)}}{\tau_{hf}}, \tag{A3}$$

$$\varepsilon^2 : \partial_{t_2} h_\alpha^{(1)} + \left(\partial_{t_1} + \boldsymbol{e}_\alpha \cdot \boldsymbol{\nabla}_1\right) h_\alpha^{(1)} + \frac{1}{\tau_h} h_\alpha^{(2)} = Z_\alpha \frac{f_\alpha^{(2)}}{\tau_{hf}}. \tag{A4}$$

**Mass conservation**

Summations of Eqs. (A1) and (A2) over $\alpha$ give, respectively

$$\partial_{t_1} \rho + \partial_{1k} (\rho u_k) = 0, \tag{A5}$$

$$\partial_{t_2} \rho = 0. \tag{A6}$$

Combining Eq. (A5) with Eq. (A6) leads to the continuum equation

$$\partial_t \rho + \boldsymbol{\nabla} \cdot (\rho \boldsymbol{u}) = 0. \tag{A7}$$

**Momentum conservation**

The first-order velocity moments of Eqs. (A1) and (A2) give, respectively

$$\partial_{t_1}(\rho u_j) + \partial_{1i}\left(\sum_\alpha e_{\alpha i} e_{\alpha j} f_\alpha^{eq}\right) = 0, \tag{A8}$$

$$\partial_{t_2}(\rho u_j) + \partial_{1i}\left(\sum_\alpha e_{\alpha i} e_{\alpha j} f_\alpha^{(1)}\right) = 0. \tag{A9}$$

According to Eq. (19a), Eq. (A8) is given by

$$\partial_{t_1}(\rho u_j) + \partial_{1k}(\rho u_k u_j) = -\partial_{1j} p, \tag{A10}$$

where $p = \rho RT$. From Eq. (A1), we can obtain

$$\sum_\alpha e_{\alpha i} e_{\alpha j} f_\alpha^{(1)} = -\tau_f \left[\partial_{t_1}\left(\sum_\alpha e_{\alpha i} e_{\alpha j} f_\alpha^{eq}\right) + \partial_{1k}\left(\sum_\alpha e_{\alpha i} e_{\alpha j} e_{\alpha k} f_\alpha^{eq}\right) - \sum_\alpha e_{\alpha i} e_{\alpha j} C_\alpha^{(0)}\right]. \tag{A11}$$

With Eqs. (19a) and (20) as well as the constraints on the correction term $C_\alpha$, Eq. (A11) can be rewritten as

$$\sum_\alpha e_{\alpha i} e_{\alpha j} f_\alpha^{(1)} = -\tau_f \left\{\partial_{t_1}(\rho u_i u_j + p\delta_{ij}) + \partial_{1k}\left[p(u_k \delta_{ij} + u_j \delta_{ik} + u_i \delta_{jk})\right]\right\}. \tag{A12}$$

Combining Eq. (A10) with Eq. (A5), we can obtain

$$\partial_{t_1}(\rho u_i u_j) = -u_i \partial_{1j} p - u_j \partial_{1i} p - \partial_{1k}(\rho u_i u_j u_k). \tag{A13}$$

In the low Mach number limit, the third-order velocity term in Eq. (A13) can be neglected. To get the expression of $\partial p/\partial t_1$, the energy conservation should be considered. The summation of Eq. (A3) gives

$$\partial_{t_1}(\rho E) + \partial_{1k}\left[(\rho E + p)u_k\right] = 0. \tag{A14}$$

The total energy $E$ in $h_\alpha^{eq}$ is defined by $E = (bRT + u^2)/2$, where $b$ is the number of degrees of freedom of a gas. Then the left-hand side of Eq. (A14) can be rewritten as

$$\partial_{t_1}\left(\frac{1}{2}\rho u^2 + \rho \frac{b}{2} RT\right) + \partial_{1k}\left(\frac{1}{2}\rho u^2 u_k + \rho \frac{b}{2} RT u_k + p u_k\right) = u_j\left[\partial_{t_1}(\rho u_j) + \partial_{1k}(\rho u_k u_j) + \partial_{1j} p\right]$$

$$+ \partial_{t_1}\left(\rho \frac{b}{2} RT\right) + \partial_{1k}\left(\rho \frac{b}{2} RT u_k\right) + p \partial_{1k} u_k. \tag{A15}$$

With the aid of Eq. (A10), we can get

$$\partial_{t_1} p = -\partial_{1k}(pu_k) - \frac{2}{b} p \partial_{1k} u_k. \tag{A16}$$

Substituting Eqs. (A13) and (A16) into Eq. (A12) gives

$$\sum_\alpha e_{\alpha i} e_{\alpha j} f_\alpha^{(1)} = -\tau_f p \left[ \left( \partial_{1i} u_j + \partial_{1j} u_i \right) - \frac{2}{b} \partial_{1k} u_k \delta_{ij} \right]. \tag{A17}$$

Combining Eqs. (A10) and (A9) with Eq. (A17), the following momentum equation can be obtained:

$$\partial_t (\rho \mathbf{u}) + \nabla \cdot (\rho \mathbf{u}\mathbf{u}) = -\nabla p + \nabla \cdot \mathbf{\Pi}, \tag{A18}$$

where $\mathbf{\Pi} = \mu \left[ \nabla \mathbf{u} + (\nabla \mathbf{u})^\mathrm{T} - (2/3)(\nabla \cdot \mathbf{u}) \mathbf{I} \right] + \mu_B (\nabla \cdot \mathbf{u}) \mathbf{I}$, $\mu = \tau_f p$, and $\mu_B = 2(1/3 - 1/b) \mu$.

**Energy conservation**

Taking summation of Eq. (A4) over $\alpha$, we can obtain

$$\partial_{t_2}(\rho E) + \partial_{1j} \left( \sum_\alpha e_{\alpha j} h_\alpha^{(1)} \right) = 0. \tag{A19}$$

Multiplying Eq. (A3) by $e_{\alpha j}$ and then taking summation over $\alpha$, we have

$$\sum_\alpha e_{\alpha j} h_\alpha^{(1)} = -\tau_h \left[ \partial_{t_1} \left[ (\rho E + p) u_j \right] + \partial_{1i} \left( \sum_\alpha e_{\alpha i} e_{\alpha j} h_\alpha^{eq} \right) \right] + \frac{\tau_h}{\tau_{hf}} u_i \sum_\alpha e_{\alpha i} e_{\alpha j} f_\alpha^{(1)}. \tag{A20}$$

Combining Eq. (A5) with Eq. (A8) leads to

$$\partial_{t_1} u_j = -u_i \partial_{1i} u_j - \frac{1}{\rho} \partial_{1j} p. \tag{A21}$$

With Eqs. (A14), (A16), and (A21), we can obtain

$$\partial_{t_1} \left[ (\rho E + p) u_j \right] = -(\rho E + p) u_i \partial_{1i} u_j - (E + RT) \partial_{1j} p - u_j \partial_{1i} \left[ (\rho E + p) u_i \right]$$
$$- u_j \partial_{1i}(p u_i) - \frac{2}{b} p u_i \delta_{ij} \partial_{1k} u_k. \tag{A22}$$

According to Eq. (19b), the following equation can be obtained after some standard algebra:

$$\partial_{t_1}\left[(\rho E + p) u_j\right] + \partial_{1i}\left( \sum_\alpha e_{\alpha i} e_{\alpha j} h_\alpha^{eq} \right) = p u_i \left( \partial_{1j} u_i + \partial_{1i} u_j \right) + (c_v + R) p \partial_{1j} T - \frac{2}{b} p u_i \delta_{ij} \partial_{1k} u_k, \tag{A23}$$

where $c_v = bR/2$. With Eqs. (A17) and (A23), we can rewrite Eq. (A19) as

$$\partial_{t_2}(\rho E) = \partial_{1j} \left[ \left( \tau_h + \frac{\tau_f \tau_h}{\tau_{hf}} \right) p u_i \left( \partial_{1j} u_i + \partial_{1i} u_j - \frac{2}{b} \delta_{ij} \partial_{1k} u_k \right) + \tau_h (c_v + R) p \partial_{1j} T \right]. \tag{A24}$$

Note that $\left( \tau_h + \tau_f \tau_h / \tau_{hf} \right) = \tau_f$ when $\tau_{hf} = \tau_h \tau_f / (\tau_f - \tau_h)$. Finally, combining Eq. (A14) with Eq.

(A24), we can obtain the macroscopic energy equation as follows:

$$\partial_t(\rho E) + \nabla \cdot [(\rho E + p)\boldsymbol{u}] = \nabla \cdot (\lambda \nabla T) + \nabla \cdot (\boldsymbol{u} \cdot \boldsymbol{\Pi}),  \tag{A25}$$

where $\lambda = \tau_h p(c_v + R)$ is the thermal conductivity.

**APPENDIX B: THE TRANSFORMATION MATRIX AND ITS INVERSE MATRIX**

$$\mathbf{M} = \begin{bmatrix} 1 & 1 & 1 & 1 & 1 & 1 & 1 & 1 & 1 \\ -4 & -1 & -1 & -1 & -1 & 2 & 2 & 2 & 2 \\ 4 & -2 & -2 & -2 & -2 & 1 & 1 & 1 & 1 \\ 0 & 1 & 0 & -1 & 0 & 1 & -1 & -1 & 1 \\ 0 & -2 & 0 & 2 & 0 & 1 & -1 & -1 & 1 \\ 0 & 0 & 1 & 0 & -1 & 1 & 1 & -1 & -1 \\ 0 & 0 & -2 & 0 & 2 & 1 & 1 & -1 & -1 \\ 0 & 1 & -1 & 1 & -1 & 0 & 0 & 0 & 0 \\ 0 & 0 & 0 & 0 & 0 & 1 & -1 & 1 & -1 \end{bmatrix}. \tag{B1}$$

$$\mathbf{M}^{-1} = \begin{bmatrix} \frac{1}{9} & -\frac{1}{9} & \frac{1}{9} & 0 & 0 & 0 & 0 & 0 & 0 \\ \frac{1}{9} & -\frac{1}{36} & -\frac{1}{18} & \frac{1}{6} & -\frac{1}{6} & 0 & 0 & \frac{1}{4} & 0 \\ \frac{1}{9} & -\frac{1}{36} & -\frac{1}{18} & 0 & 0 & \frac{1}{6} & -\frac{1}{6} & -\frac{1}{4} & 0 \\ \frac{1}{9} & -\frac{1}{36} & -\frac{1}{18} & -\frac{1}{6} & \frac{1}{6} & 0 & 0 & \frac{1}{4} & 0 \\ \frac{1}{9} & -\frac{1}{36} & -\frac{1}{18} & 0 & 0 & -\frac{1}{6} & \frac{1}{6} & -\frac{1}{4} & 0 \\ \frac{1}{9} & \frac{1}{18} & \frac{1}{36} & \frac{1}{6} & \frac{1}{12} & \frac{1}{6} & \frac{1}{12} & 0 & \frac{1}{4} \\ \frac{1}{9} & \frac{1}{18} & \frac{1}{36} & -\frac{1}{6} & -\frac{1}{12} & \frac{1}{6} & \frac{1}{12} & 0 & -\frac{1}{4} \\ \frac{1}{9} & \frac{1}{18} & \frac{1}{36} & -\frac{1}{6} & -\frac{1}{12} & -\frac{1}{6} & -\frac{1}{12} & 0 & \frac{1}{4} \\ \frac{1}{9} & \frac{1}{18} & \frac{1}{36} & \frac{1}{6} & \frac{1}{12} & -\frac{1}{6} & -\frac{1}{12} & 0 & -\frac{1}{4} \end{bmatrix}. \tag{B2}$$

**REFERENCES**

[1] U. Frisch, B. Hasslacher, and Y. Pomeau, Phys. Rev. Lett., **56**, 1505 (1986).

[2] R. Benzi, S. Succi, and M. Vergassola, Phys. Rep., **222**, 145 (1992).

[3] S. Chen and G. D. Doolen, Annu. Rev. Fluid Mech., **30**, 329 (1998).

[4] S. Succi, *The Lattice Boltzmann Equation for Fluid Dynamics and Beyond* (Clarendon Press, Oxford, 2001).

[5] H. Lai and C. Ma, J. Stat. Mech. P04011, (2010); H. Lai and C. Ma, Physica A, **388**, 1405 (2009).

[6] S. Succi, Eur. Phys. J. B **64**, 471 (2008).

[7] J. M. V. A. Koelman, Europhys. Lett., **15**, 603 (1991).

[8] S. Chen, H. Chen, D. Martínez, and W. Matthaeus, Phys. Rev. Lett. **67**, 3776 (1991).

[9] Y. H. Qian, D. d'Humières, and P. Lallemand, Europhys. Lett. **17**, 479 (1992).

[10] X. He and L.-S. Luo, Phys. Rev. E **56**, 6811 (1997).

[11] D. d'Humières, in *Rarefied Gas Dynamics: Theory and Simulations*, Prog. Astronaut. Aeronaut. Vol. 159, edited by B. D. Shizgal and D. P. Weaver (AIAA, Washington, D.C., 1992).

[12] P. Lallemand and L.-S. Luo, Phys. Rev. E **61**, 6546 (2000).

[13] D. d'Humières, I. Ginzburg, M. Krafczyk, P. Lallemand, and L.-S. Luo, Phil. Trans. R. Soc. Lond. A **360**, 437 (2002).

[14] L.-S. Luo, W. Liao, X. Chen, Y. Peng, and W. Zhang, Phys. Rev. E **83**, 056710 (2011).

[15] Q. Li, Y. L. He, G. H. Tang, and W. Q. Tao, Phys. Rev. E **81**, 056707 (2010).

[16] F. Chen, A. Xu, G. Zhang, Y. Li, and S. Succi, Europhys. Lett. **90**, 54003 (2010)

[17] I. Ginzburg, Adv. Water Resour. **28**, 1196 (2005).

[18] I. Ginzburg, Comm. Comp. Phys. **3**, 427 (2008).

[19] D. O. Martínez, W. H. Matthaeus, S. Chen, and D. C. Montgomery, Phys. Fluids **6**, 1285 (1994).

[20] F. J. Alexander, S. Chen, and J. D. Sterling, Phys. Rev. E **47**, R2249 (1993).

[21] Y. H. Qian, J. Sci. Comput. **8**, 231 (1993).

[22] Y. Chen, H. Ohashi, and M. Akiyama, Phys. Rev. E **50**, 2776 (1994).

[23] X. Shan, X.-F. Yuan, and H. Chen, J. Fluid Mech. **550**, 413 (2006).

[24] S. S. Chikatamarla, I. V. Karlin, Phys. Rev. Lett. **97**, 190601 (2006); Comput. Phys. Commun. **179**, 140 (2008).

[25] Y. Gan, A. Xu, G. Zhang, X. Yu, and Y. Li, Physica A **387**, 1721 (2008).

[26] Y. Gan, A. Xu, G. Zhang, and Y. Li, Phys. Rev. E **83**, 056704 (2011).

[27] X. Shan, Phys. Rev. E **55**, 2780 (1997).

[28] X. He, S. Chen, and G. D. Doolen, J. Comput. Phys. **146**, 282 (1998).

[29] Y. Peng, C. Shu, and Y. T. Chew, Phys. Rev. E **68**, 026701 (2003).

[30] Y. Shi, T. S. Zhao, and Z. L. Guo, Phys. Rev. E **70**, 066310 (2004).

[31] Z. Guo, C. Zheng, B. Shi, and T. S. Zhao, Phys. Rev. E **75**, 036704 (2007).

[32] S. Chen, J. Tölke, and M. Krafczyk, Phys. Rev. E **79**, 016704 (2009).

[33] Q. Li, Y. L. He, Y. Wang, and W. Q. Tao, Phys. Rev. E **76**, 056705 (2007).

[34] Q. Li, Y. L. He, G. H. Tang, and W. Q. Tao, Phys. Rev. E **80**, 037702 (2009).

[35] P. Lallemand and L.-S. Luo, Phys. Rev. E **68**, 036706 (2003).

[36] P. Lallemand and L.-S. Luo, Int. J. Mod. Phys. C **17**, 41 (2003).

[37] N. I. Prasianakis and I. V. Karlin, Phys. Rev. E **76**, 016702 (2007).

[38] N. I. Prasianakis and I. V. Karlin, Phys. Rev. E **78**, 016704 (2008).

[39] P. C. Philippi, L. A. Hegele, L. O. E. dos Santos, and R. Surmas, Phys. Rev. E **73**, 056702 (2006).

[40] Z. L. Guo and C. G. Zheng, *Theory and Applications of Lattice Boltzmann Method* (Science Press, Beijing, 2009).

[41] M. E. MaCracken and J. Abraham, Phys. Rev. E **71**, 036701 (2005).

[42] T. Lee and P. F. Fischer, Phys. Rev. E **74**, 046709 (2006).

[43] C. M. Pooley and K. Furtado, Phys. Rev. E **77**, 046702 (2008).

[44] H. W. Liepmann and A. Roshko, *Elements of Gasdynamics* (Wiley, New York, 1957).

[45] Z. L. Guo, C. Zheng, and B. Shi, Chin. Phys. **11**, 0366 (2002).

[46] D. Haydock and J. M. Yeomans, J. Phys. A: Math. Gen. **36**, 5683 (2003).

[47] G. Emanuel, Int. J. Eng. Sci. **36**, 1313 (1998).

[48] M. Hortmann, M. Perić, and G. Scheuerer, Int. J. Numer. Methods Fluids **11**, 189 (1990).

[49] J. Vierendeels, B. Merci, and E. Dick, Int. J. Numer. Meth. Heat Fluid Flow **13**, 1057 (2003).

[50] G. H. Tang, P. X. Ye, and W. Q. Tao, J. Non-Newtonian Fluid Mech. **165**, 435 (2010).

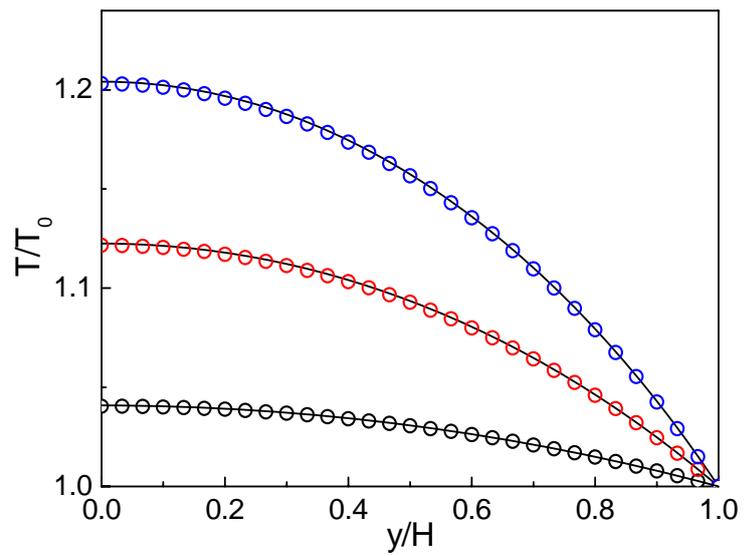

FIG. 1. Dimensionless temperature profiles of thermal Couette flow at $\gamma = 5/3$ and $\mathrm{Ma} = 0.35$. From top to bottom: $\mathrm{Pr} = 5$, $3$, and $1$. Solid lines and the circles are the analytical and numerical results, respectively.

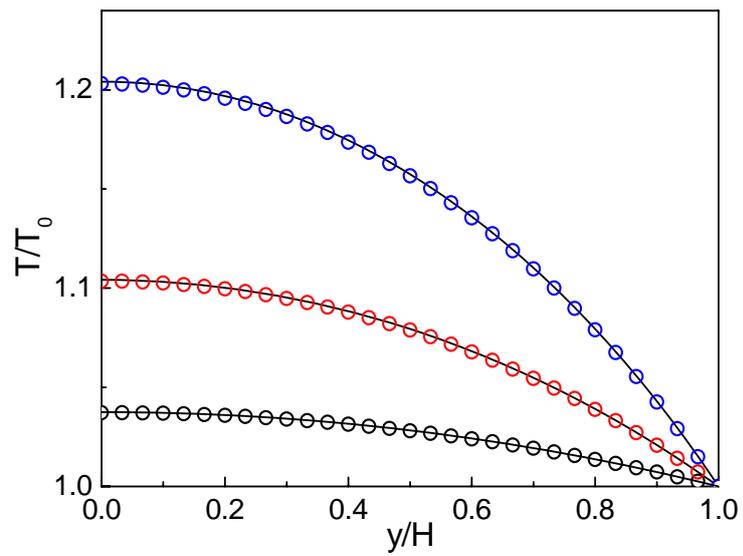

FIG. 2. Dimensionless temperature profiles of thermal Couette flow at $\gamma = 5/3$ and $Pr = 5$. From top to bottom: $Ma = 0.35$, $0.25$, and $0.15$. Solid lines and the circles are the analytical and numerical results, respectively.

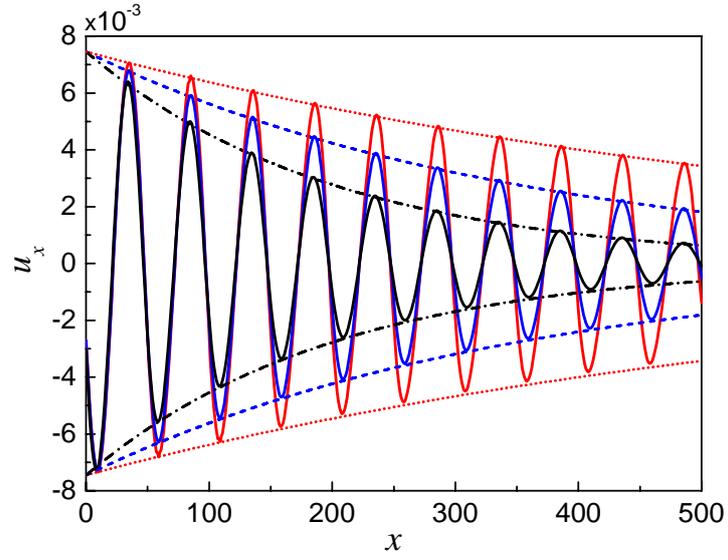

FIG. 3. Velocity profiles of attenuation-driven acoustic streaming at $\gamma = 5/3$ with different Prandtl numbers. The solid lines are the numerical results. The dotted lines ($Pr = 5$), dashed lines ($Pr = 0.5$), and dashdotted lines ($Pr = 0.2$) are the analytical velocity amplitude (see Eq. (55) and note that $\pm U(x)$ are plotted).

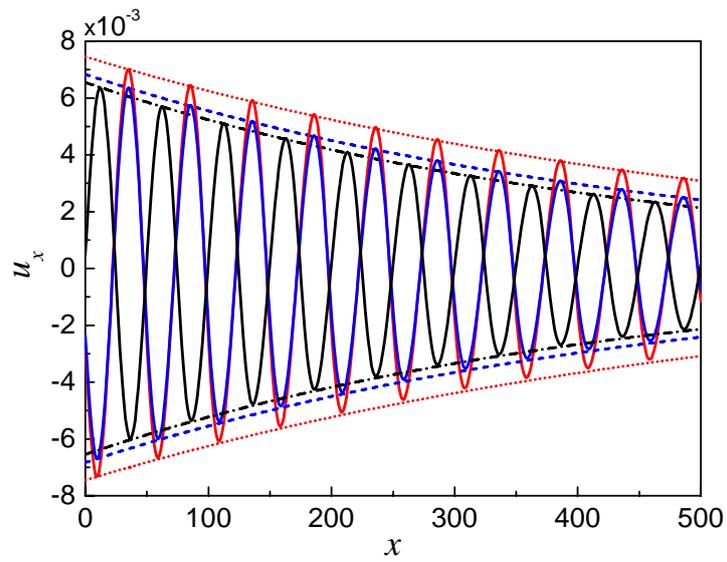

FIG. 4. Velocity profiles of attenuation-driven acoustic streaming at Pr = 2 with different specific-heat ratios. The solid lines are the numerical results. The dotted lines ($\gamma = 5/3$), dashed lines ($\gamma = 7/5$), and dashdotted lines ($\gamma = 9/7$) are the analytical velocity amplitude.

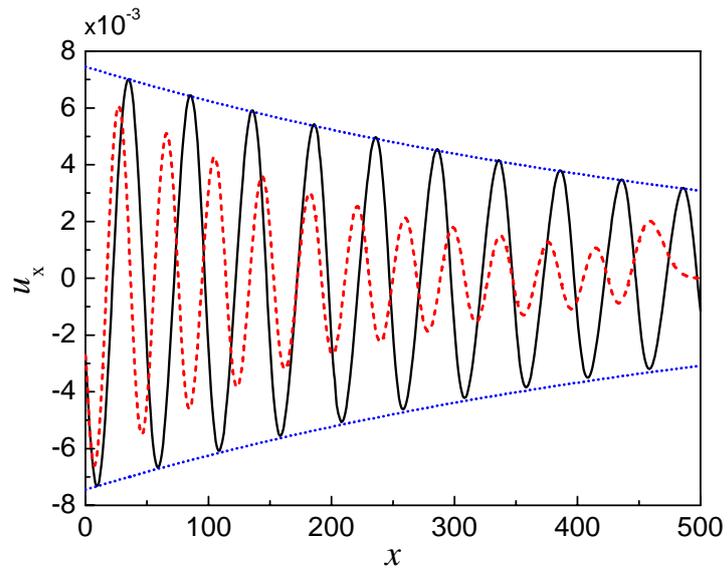

FIG. 5. Comparison of the velocity profiles of attenuation-driven acoustic streaming at $Pr = 2$ and $\gamma = 5/3$ between the present model (the solid line) and the decoupling model (the dashed line). The dotted lines are the analytical velocity amplitude.

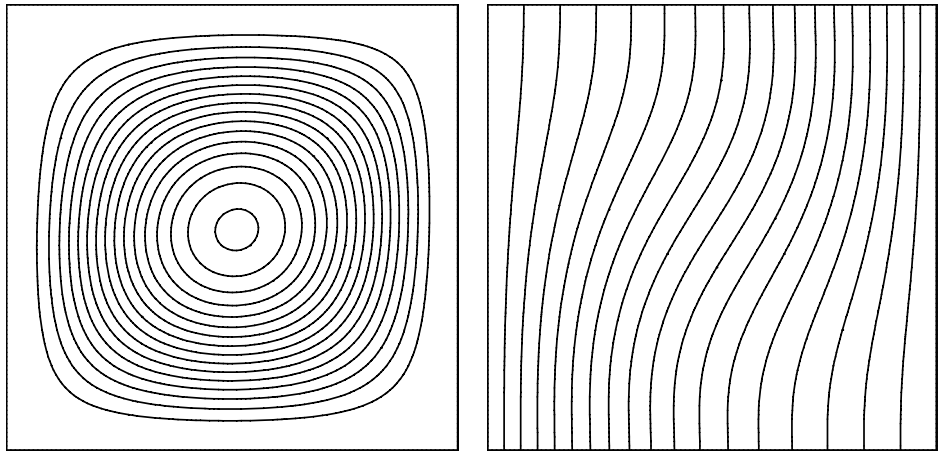

(a) $Ra = 10^3$

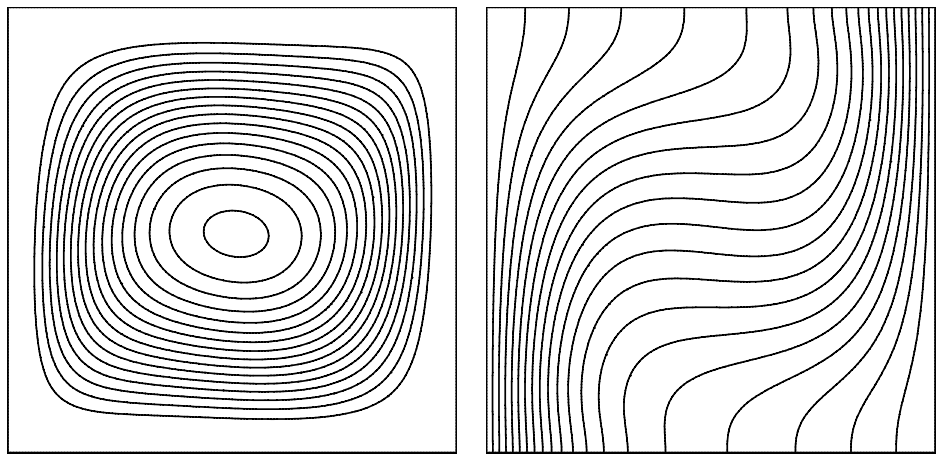

(b) $Ra = 10^4$

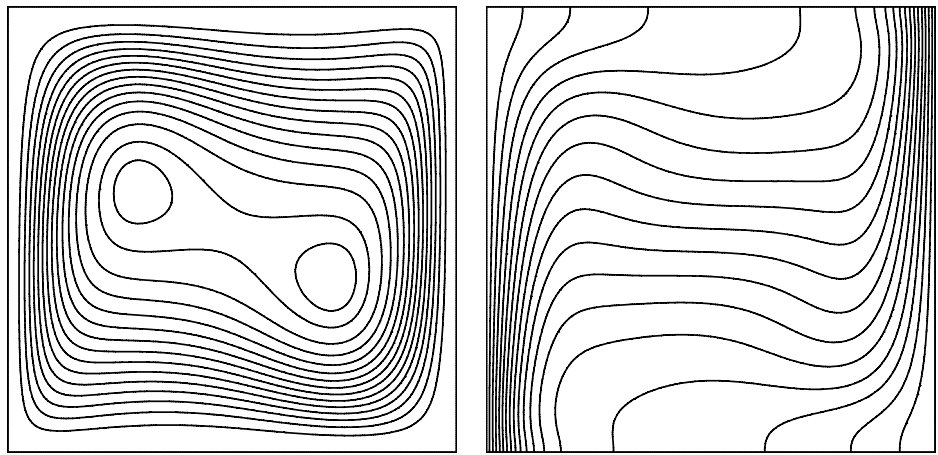

(c) $Ra = 10^5$

FIG. 6. Streamlines (left) and isotherms (right) of natural convection in a square cavity at

$Ra = 10^3, 10^4,$ and $10^5$ with $T_h = 315\,\text{K}$ and $T_l = 285\,\text{K}$.

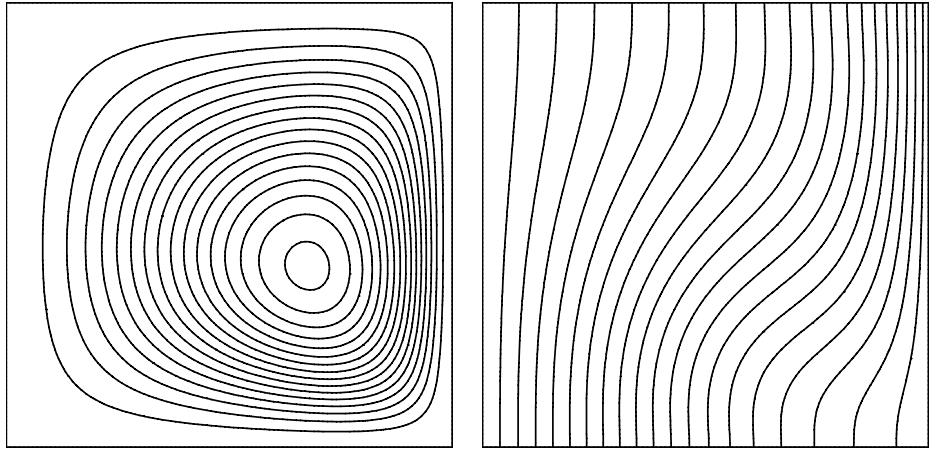

(a) $Ra = 10^3$

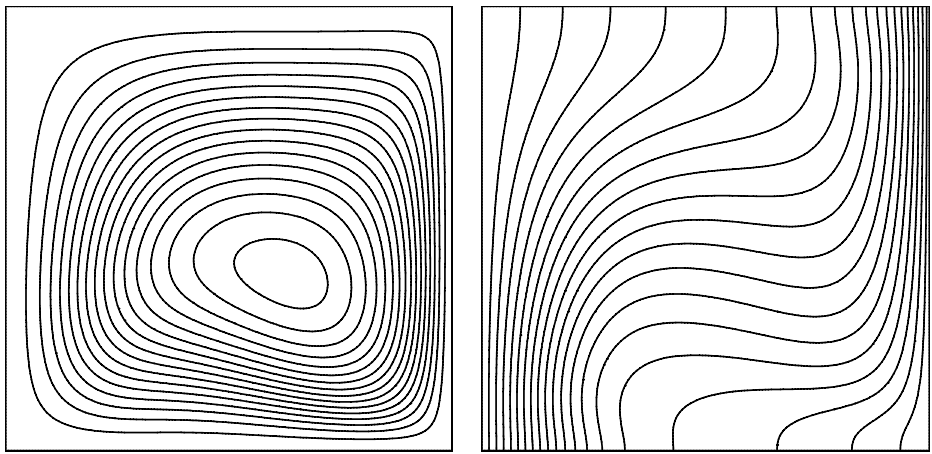

(b) $Ra = 10^4$

FIG. 7. Streamlines (left) and isotherms (right) of natural convection in a square cavity at $Ra = 10^3$ and $10^4$ with $T_h = 960\,\text{K}$ and $T_l = 240\,\text{K}$.

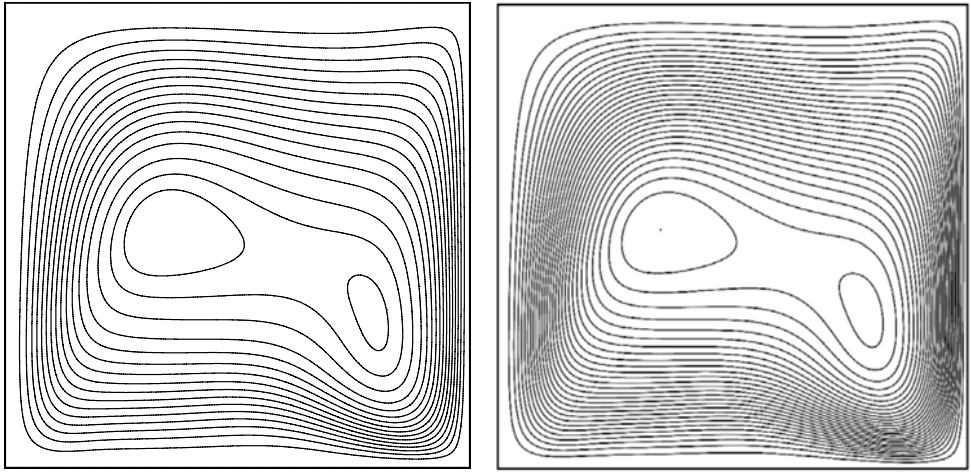

(a) streamlines

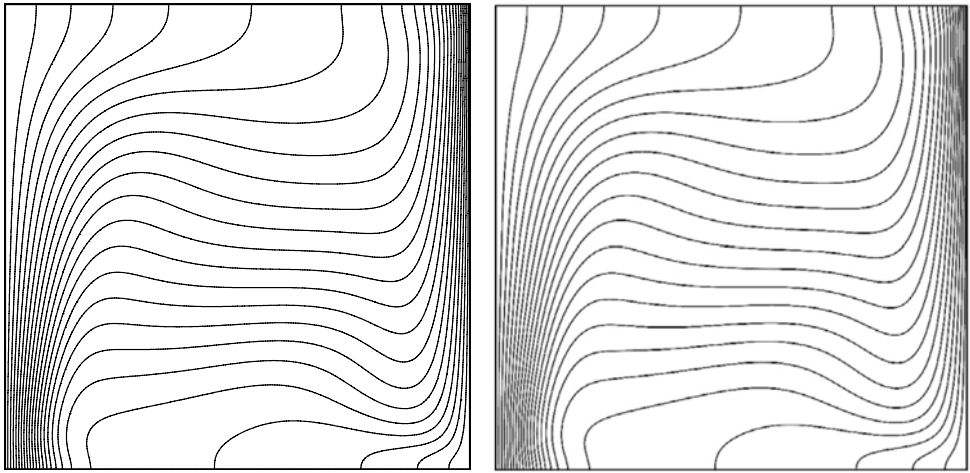

(b) isotherms

FIG. 8. Comparions of streamlines and isotherms of natural convection in a square cavity at

$\text{Ra} = 10^5$ with $T_h = 960\,\text{K}$ and $T_l = 240\,\text{K}$ : Present (left) and Ref. [49] (right).

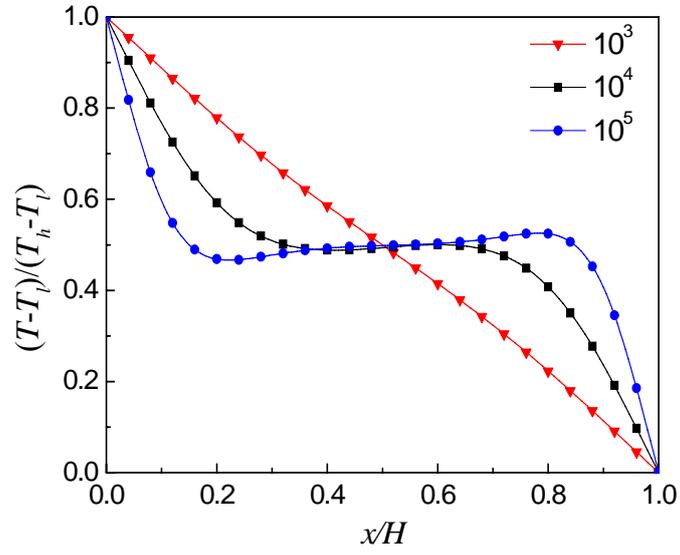

(a)

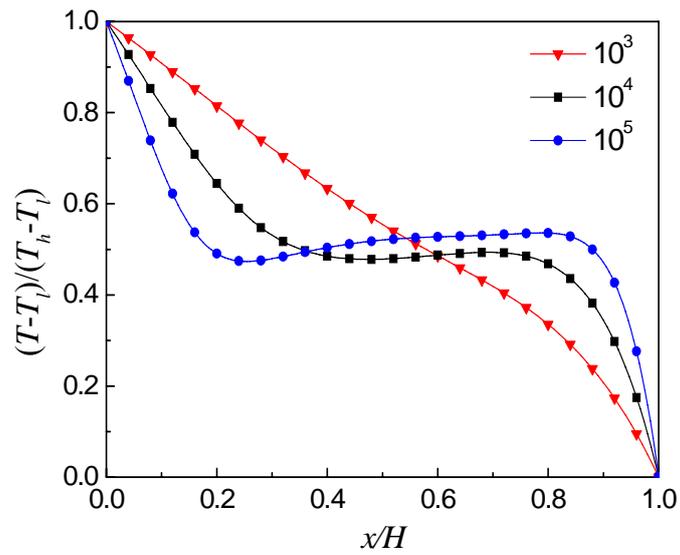

(b)

FIG. 9. Comparions of the temperature profiles along the horizontal line crossing the center of the cavity: (a) $T_h = 315\,\text{K}$ and $T_l = 285\,\text{K}$; and (b) $T_h = 960\,\text{K}$ and $T_l = 240\,\text{K}$.

Table I. Comparison of the numerical ($T$) and analytical ($T_a$) temperatures at the lower wall of thermal Couette flow in different cases.

| Cases | Pr = 1, Ma = 0.35 | Pr = 3, Ma = 0.35 | Pr = 5, Ma = 0.35 | Pr = 5, Ma = 0.25 | Pr = 5, Ma = 0.15 |
|---|---|---|---|---|---|
| $T/T_0$ | 1.0405 | 1.1217 | 1.2032 | 1.1036 | 1.0373 |
| $T_a/T_0$ | 1.0408 | 1.1225 | 1.2042 | 1.1042 | 1.0375 |
| $E_r$ (%) | 0.74 | 0.65 | 0.49 | 0.58 | 0.53 |

Table II. Comparisons of the average Nusselt number and the maximum velocity components across the cavity center.

| Ra | | Nu | $u_{max}$ | $y_{max}$ | $v_{max}$ | $x_{max}$ |
|---|---|---|---|---|---|---|
| $10^3$ | Present | 1.1207 | 3.664 | 0.8100 | 3.699 | 0.1800 |
| | Ref. [48] | - | 3.649 | 0.8130 | 3.697 | 0.1780 |
| | Ref. [31] | 1.1195 | 3.643 | 0.8047 | 3.692 | 0.1719 |
| $10^4$ | Present | 2.2528 | 16.351 | 0.8200 | 19.589 | 0.1200 |
| | Ref. [48] | 2.2448 | 16.180 | 0.8265 | 19.630 | 0.1193 |
| | Ref. [31] | 2.2545 | 16.125 | 0.8203 | 19.558 | 0.1172 |
| $10^5$ | Present | 4.5350 | 35.703 | 0.8550 | 68.536 | 0.0650 |
| | Ref. [48] | 4.5216 | 34.740 | 0.8558 | 68.640 | 0.0657 |
| | Ref. [31] | 4.5278 | 34.603 | 0.8516 | 68.082 | 0.0703 |